\documentclass[structabstract]{aa}
\usepackage{txfonts}
\usepackage{graphicx}
\usepackage{natbib}
\usepackage{hyperref}
\bibpunct{(}{)}{;}{a}{}{,} % to follow the A&A style

\begin{document}
\title{A spectroscopic survey of thick disc stars \\
outside the solar neighbourhood \thanks{Based on VLT/FLAMES observations collected at the European Southern Observatory, proposals 075.B-0610-A \& 077.B-0.382.}\fnmsep \thanks{Tables~\ref{online3}, ~\ref{online} and  ~\ref{online2} 
are only available in electronic form at the CDS via anonymous ftp to cdsarc.u-strasbg.fr (130.79.128.5)
or via \url{http://cdsweb.u-strasbg.fr/cgi-bin/qcat?J/A+A/}}}
\author{G.~Kordopatis \inst{\ref{inst1}}
  \and A.~Recio-Blanco \inst{\ref{inst1}}
  \and P.~de~Laverny \inst{\ref{inst1}}
  \and G.~Gilmore \inst{\ref{inst2}}
  \and V.~Hill \inst{\ref{inst1}}
  \and R.F.G.~Wyse \inst{\ref{inst3}}
  \and A.~Helmi \inst{\ref{inst4}}
  \and A.~Bijaoui \inst{\ref{inst1}}
  %\and C.~Ordenovic \inst{\ref{inst1}}
  \and M.~Zoccali \inst{\ref{inst5}}
  \and O.~Bienaym\'e \inst{\ref{inst6}}}
\institute{Universit\'e de Nice Sophia Antipolis, CNRS, Observatoire de la C\^ote d'Azur, Cassiop\'ee UMR 6202, BP 4229, 06304 Nice, France  \email{georges.kordopatis@oca.eu} \label{inst1}
\and Institute of Astronomy, University of Cambridge, Madingley Road, Cambridge CB3 0HA, UK \label{inst2}
\and Johns Hopkins University, Baltimore, MD, USA  \label{inst3}
\and Kapteyn Astronomical Institute, University of Groningen, PO Box 800, 9700 AV Groningen, The Netherlands \label{inst4}
\and Departamento de Astronom\'ia y Astrof\'isica, Pontificia Universidad Cat\'olica de Chile, Av. Vicu\~na Mackenna 4860, Casilla 306, Santiago 22, Chile \label{inst5}
\and Universit\'e de Strasbourg, Observatoire Astronomique, Strasbourg, France \label{inst6} }
\date{Received / Accepted}

\newcommand{\VphiMetaCorrel}{$\partial V_\phi /\partial [M/H]=-45 \pm 12$~km~s~$^{-1}$~dex~$^{-1}$}
\newcommand{\VphiZ}{$\partial V_\phi /\partial Z=19 \pm 8$~km~s~$^{-1}$~kpc~$^{-1}$}
\newcommand{\MetaZ}{$\partial [M/H] /\partial Z=-0.14 \pm 0.05$~dex~kpc~$^{-1}$}
\newcommand{\SigmaVphiZ}{$\partial \sigma_{V_\phi}/ \partial Z=9 \pm 7$~\mbox{km s}$^{-1}$\mbox{kpc}$^{-1}$}
\newcommand{\HrValue}{$h_R\sim3.4 \pm 0.7$~kpc}
\newcommand{\HzValue}{$h_Z\sim694 \pm 45$~pc}

% \abstract{}{}{}{}{} 
% 5 {} token are mandatory
 
  \abstract
   { In the era of large spectroscopic surveys, galactic archaeology aims to understand the formation and evolution of the Milky Way by means of large datasets. In particular, the kinematic and chemical study of the thick disc can give valuable information on the merging history of the Milky Way.}
   {Our aim is to detect and characterise  the galactic thick disc chemically and dynamically  by analysing F, G and K stars, whose atmospheres reflect their initial chemical composition.}
   { We performed a spectroscopic survey of nearly 700 stars probing the galactic thick disc far from the solar neighbourhood towards the galactic coordinates ($l\sim 277^{\circ}$, $b\sim 47^{\circ}$).  The derived effective temperatures, surface gravities and overall metallicities were then combined with stellar evolution isochrones, radial velocities and 
     proper motions to derive the distances, kinematics and  orbital parameters of the sample stars. The targets belonging to each galactic  component (thin disc, thick disc, halo) were selected either on their kinematics  or according to their position above the galactic plane, and the vertical gradients were also estimated.}
   { We present here atmospheric parameters, distances and kinematics for this sample, and a comparison of our kinematic and metallicity 
     distributions with the Besan\c{c}on model of the Milky Way. 
     The thick disc far from the solar neighbourhood is found to differ only slightly from the thick disc properties as derived in the solar vicinity. For regions where the thick disc dominates ($1\lesssim Z \lesssim 4$~kpc), we measured vertical velocity and metallicity trends of  \VphiZ\ and \MetaZ, respectively.  These trends can be explained as a smooth transition between the different galactic components, although intrinsic gradients could not be excluded. In addition, a correlation \VphiMetaCorrel \ between the orbital velocity and the metallicity of the thick disc is detected. This gradient is inconsistent with the SDSS photometric survey analysis, which did not detect any such trend, and challenges
radial migration models of thick disc formation. Estimations of the scale heights and scale lengths 
     for different metallicity bins of the thick disc result in consistent values, with \HrValue, and \HzValue, showing no evidence of 
     relics of destroyed massive satellites. }{}


\maketitle

%
%________________________________________________________________

\section{Introduction}
\label{sec:introduction}

In the paradigm of a dark energy and dark matter dominated Universe, the process of disc galaxy formation is still
 very poorly understood. For instance, the creation of the galactic thick disc still remains a riddle. 
Was it formed by the accretion of satellites that have deposited their debris in a roughly planar configuration
\citep{Galaxy_model_Abadi}, or was the thin disc heated after successive small accretions as suggested for example  by 
\citet{Galaxy_model_Villalobos}? Did the thick disc stars form {\it in situ}, through heating of the thin disc by 
gas rich mergers and starburst during the merger process \citep{Galaxy_model_Brook}, or did they migrate because of  
resonances with the spiral arms (and the central bar), as suggested for example by \citet{Radial_mixing_2009} or \citet{Roskar_2008}? 

To answer these questions, we would ideally like to tag or associate the visible components of the Galaxy to 
parts of the proto-galactic hierarchy. All necessary constraints can be obtained from a detailed analysis of the 
chemical abundances of cool and intermediate temperature stars. Indeed, F, G and K type dwarf stars are 
particularly useful to study galactic evolution, because they are both numerous and long-lived, and their 
atmospheres reflect their initial chemical 
composition. However, a direct measurement of their spatial distribution requires accurate estimates of stellar 
 distances, which is a delicate step involving (if the parallax is not available) the determination of precise 
stellar parameters (effective temperatures, surface gravities and metal content).

That is the reason why most spectroscopic surveys of the thick disc up to now  are restricted to the solar 
neighbourhood (within about a 500~pc radius) except for those with relatively few stars in their samples. Nevertheless, 
detailed metallicity measurements based on such spectroscopic observations of kinematically selected thick disc 
stars revealed in several studies  a clear distinction in chemical elements ratios between the metal-poor tails 
of the thin and the thick discs \citep{Edvardsson_93, Bensby_2006, Reddy_2006, Fuhrmann_2008, Ruchti_2010, Navarro11}. 
Stars belonging to the latter are $\alpha$-enhanced, suggesting a rapid formation of the thick disc and a distinct 
chemical history. Indeed, the [$\alpha$/Fe] chemical index is commonly used to trace the star formation time-scale 
in a system in terms of the distinct roles played by supernovae (SNe) of different types in the galactic enrichment. 
The $\alpha$-elements are produced mainly during Type II SNe explosions of massive stars (M $> 8 M_\odot$) 
on a short time-scale ($\sim 10^7$ years), whereas iron is also produced by Type Ia SNe of less massive stars 
on a much longer time-scale ($\sim  10^9$ years).  

On the other hand, photometric surveys such as the SDSS \citep{SDSS_description} explore a much larger volume, but suffer from greater 
uncertainties in the derived parameters. Based  on photometric metallicities and distances of more than 2 million F/G stars up to 
8~kpc from the Sun, \citet{Ivezic_2008} suggested that the transition between the thin and the thick disc can be modelled as smooth, 
vertical shifts of metallicity and velocity distributions, challenging the view of two distinct populations that was  introduced by \citet{Gilmore_1983}. 

The goal of this paper is to put additional constraints on the vertical properties of the thick disc. 
For that purpose, we spectroscopically explored the stellar contents 
outside the solar neighbourhood using the \citet{Ojha_photometry} catalogue. The authors 
of that survey photometrically observed several thousands of stars towards the direction of galactic antirotation. 
They provide the Johnson  U, B, V magnitudes as well as the proper motions for most of the targets up to V$\sim$18.5~mag. 
We selected 689 of them  based on their magnitudes to probe the galactic thick disc, and observed them 
using the  VLT/FLAMES GIRAFFE spectrograph in the LR08 setup (covering the infrared ionized calcium triplet).

 The line-of-sight ($l\sim277^{\circ}$, $b\sim47^{\circ}$) was chosen in accordance to results found by \citet{Gilmore_2002} 
towards ($l\sim270^{\circ}$, $b\sim-45^{\circ}$) and ($l\sim270^{\circ}$, $b\sim33^{\circ}$) that were confirmed by 
\citet{Wyse_2006} towards ($l\sim260^{\circ}$, $b\sim-23^{\circ}$), ($l\sim104^{\circ}$, $b\sim45^{\circ}$), 
($l\sim86^{\circ}$, $b\sim35^{\circ}$), which state that thick disc stars farther than 2~kpc from the Sun seem to have a rotational 
lag greater than the canonical thick disc. Indeed, it is commonly accepted that the latter lags the Local 
Standard of Rest (LSR) by  $\sim50$~\mbox{km s}$^{-1}$, whereas the authors cited above found a lag twice as high  
($\sim100$~\mbox{km s}$^{-1}$) at long distances.  Because the angular momentum is essentially a conserved quantity in galaxy formation, 
this lagging sub-population has been suggested by the authors to be the remnants of the last major merger of the 
Milky Way, back to  z~$\sim$~2. 

For the analysis of this very substantial sample of FLAMES spectra we used the pipeline presented in \citet[hereafter paper~I]{Kordopatis11a}.
 It allowed us to obtain the effective temperature ($T_\mathrm{eff}$), the surface gravity ($\log~g$) 
and the overall metallicity\footnote{We define the stellar overall metallicity as 
[M/H]=$\log (\frac{N(M)}{N(H)})_* - \log (\frac{N(M)}{N(H)})_{\sun}$, where $N$ represents the number density and $M$ 
all elements heavier than \ion{He}{}.} ([M/H]) for the stars of our sample. They 
were combined with the proper motions and the (B-V) colours of the Ojha catalogue and our derived radial velocities. 
Those parameters were then used to estimate the distances, galactocentric positions and kinematics, as well as the orbital 
parameters (eccentricities, apocentric and pericentric distances, angular momenta) of our targets.

The structure of this paper is as follows. The observed stellar sample is presented in the next section together with the 
data reduction and the radial velocity derivation.  The method used to determine the atmospheric parameters is reviewed 
in Sect.~\ref{sec:pipeline_atm}. In Sect.~\ref{sec:dynamics} the estimates for the atmospheric parameters, the distances and 
the kinematics are presented.   
In Sect.~\ref{subsec:star_selection} a clean sample of stars is selected, whose stars are analysed in  Sect.~\ref{sec:results} 
with respect to their distance above the galactic plane; then we compare them to the Besan\c{c}on 
model of the Milky Way \citep{Modele_besancon}. The galactic components, chosen with a probabilistic approach and according 
to their distance above the galactic plane,  are also characterised and discussed in relation with thick disc formation scenarios. 
Finally, we estimate in Sect.~\ref{sec:scale_lengths} the radial scale lengths and scale heights of the thin disc and the thick disc. 

%
%________________________________________________________________

\section{Observations, data reduction and radial velocity derivation}
\label{sec:observations}

The observations were obtained with VLT/FLAMES feeding the GIRAFFE spectrograph in MEDUSA mode, that allows a simultaneous 
allocation of 132 fibres (including sky). The GIRAFFE low-resolution grating LR08 (8206-9400~\AA, $R\sim$6500, sampling=0.2~\AA) 
was used during ESO observing periods 75 and 77 (2005 and 2006, respectively). One of the interesting points of that 
configuration is that it contains  the Gaia/RVS wavelength range (8475-8745~\AA), and is similar to its 
low-resolution mode ($R\sim$7000). 

In that wavelength range the IR \ion{Ca}{ii} triplet (8498.02, 8542.09, 8662.14~\AA) is predominant for 
most spectral types and luminosity classes even for very metal-poor stars \citep[see, for example][]{Gaia_Range}. 
In addition, these strong features are still detectable even at low signal-to-noise ratio (S/N), allowing a good radial 
velocity ($V_\mathrm{rad}$) derivation and an overall metallicity estimation. Paschen lines (for example 8502.5, 8545.4, 
8598.4, 8665.0~\AA) are visible for stars hotter than G3. The \ion{Mg}{i} (8807~\AA) line, which is a useful indicator 
of surface gravity \citep[see][]{MgI_line}, is also visible even for low S/N. Finally, molecular 
lines such as TiO and CN can be seen for cooler stars.

 The 689 stars of our sample were selected only from their V-magnitudes and the availability of  proper motions measurements in the 
catalogue of \cite{Ojha_photometry}. They were faint enough to probe the galactic thick disc and bright enough to have acceptable 
S/N ($ m_V \leq 18.5$~mag).  To fill all available fibres of MEDUSA, brighter stars in the Ojha catalogue were added to 
our survey ($m_V \geq 14$), resulting in a bimodal distribution in magnitudes.  
We made no colour selection, which resulted in a wide range of both $\log~g$ and  $T_\mathrm{eff}$ as we 
will show in Sect.~\ref{sec:pipeline_atm} and Sect.~\ref{sec:dynamics}. 
The magnitude precisions range from 0.02~mag for the brightest, to 0.05~mag for the faintest 
stars \citep{Ojha_photometry}. Associated errors for the proper motions are estimated to be 2~mas~year$^{-1}$.

Eight different fields were observed with two exposures each. The spectra were reduced using the Gasgano ESO GIRAFFE 
pipeline\footnote{\url{http://www.eso.org/gasgano}} \citep[version 2.3.0,][]{gasgano} with optimal extraction. The sky-lines were removed 
from our spectra using the software developed by  M.~Irwin \citep{Battaglia_sky_sub}, which is optimal for the wavelength range 
around the IR \ion{Ca}{ii} triplet.
For each field the four fibres that were allocated to the sky were combined to obtain a median sky spectrum and separate 
the continuum from the sky-line 
components. 
Then,  the latter was cross-correlated to the object spectrum (which was also split into continuum and line components) to match the object sky-line intensities and positions. 
Afterwards, the sky was subtracted by finding the optimal scale factor between the masked sky-line and the object spectrum.
Individual exposures were then cross-correlated  using IRAF to put them on the same reference frame and to be summed. 
Before summing them, the cosmic rays and the remaining residual sky-lines were removed by comparing the flux levels of the individual 
exposures.

\begin{figure}
  \resizebox{\hsize}{!}{\includegraphics{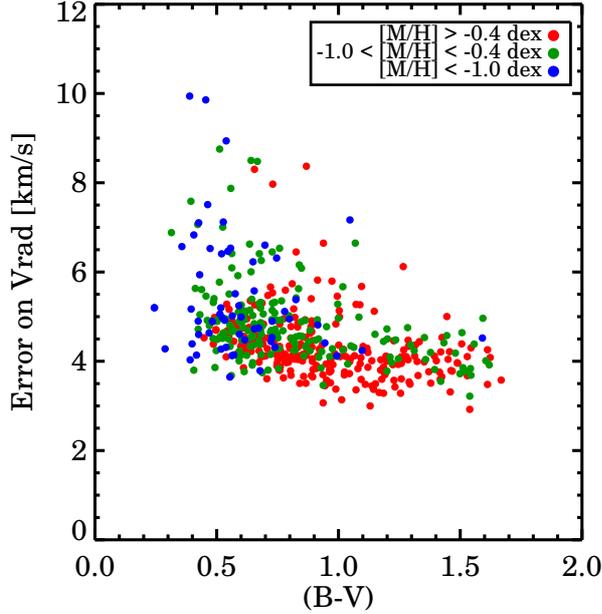}}
  \caption{Radial velocity uncertainties versus the (B-V) colour for the final catalogue  (479 stars), as selected in Sect.~\ref{subsec:star_selection}. Dots with different colours refer to different metallicity ranges.}
  \label{fig:Vrad_errors}
\end{figure}

To measure the radial velocities we used a binary mask of a K0 type star at the LR08 resolution 
\citep[available from the Geneva observatory with the girBLDRS routine, see][]{Royer_CCF}. The cross-correlation function
(CCF) between the observed spectra and the binary mask was computed, extracting from the position of the peak the 
$V_\mathrm{rad}$ value. The spectral features corresponding to the full-width at half-maximum (FWHM) of the \ion{Ca}{ii} 
triplet and the \ion{Mg}{i} (8807\AA) lines were added manually into this binary mask, because we found that otherwise the 
cross-correlation routine did not converge for the lowest  S/N spectra (where the spectral lines are hard to identify).
The  wider lines added to the binary mask play a dominant role in the cross-correlation method. This 
implies that no particular concern has to be raised about errors caused by possible template mismatches. The only expected effect
is a decrease in the precision of the $V_\mathrm{rad}$ estimates, owing to a broader CCF. 
Nevertheless, as shown in Paper~I, the derivation of the atmospheric parameters is not altered as long as the 
uncertainties in $V_\mathrm{rad}$ are less then $\sim$7-8~\mbox{km s}$^{-1}$. We found in our survey that the mean error on the 
$V_\mathrm{rad}$ is 4.70~\mbox{km s}$^{-1}$ and the standard deviation of the error distribution is 1.3~\mbox{km s}$^{-1}$.
In Fig.~\ref{fig:Vrad_errors} we plot the estimated errors on the radial velocities versus the (B-V) colour of the targets. 
A colour code was added according to the derived metallicity of the stars (as found in Sect.~\ref{sec:pipeline_atm}). The largest 
errors are found for the hotter and the most metal-poor stars, as expected, because 
 the mixture of broader lines  for the former and weaker lines for the latter differ the most.
 The individual values of the heliocentric $V_\mathrm{rad}$ are presented in the online Table~\ref{online3}, and will be used in 
Sect.~\ref{subsec:kinematics} to determine the galactocentric velocities and hence the orbits of the stars. 
%Let us note, finally, that the FWHM of the CCF can give a rough idea of the projected rotational velocity of the star.
%The pipeline of Paper~I not taking into account this effect, it was important to check our sample did not include any fast rotator.

The spectra were then shifted at the rest frame and linearly rebinned by a factor of two, in agreement with Shannon's criterion. 
Furthermore, they were resampled to match the sampling of the synthetic spectra library we had in our possession, which we used to derive
the atmospheric parameters (see Paper~I). 
This led to a final sampling of 0.4~\AA \ and an increased S/N per pixel. 

Then the spectra were cut at the wavelengths 8400-8820~\AA \ to keep the range with the predominant lines and 
suffering least by CCD spurious effects (border effects, presence of a glow in the red part) and possible sky residuals. 
For this reason, the range between 8775 and 8801~\AA \ which contains few iron lines compared to the possible important sky 
residuals, was also removed. The spectral feature corresponding to the \ion{Mg}{i} around $\sim$8807 \AA ~was nevertheless kept. 
The cores of the strong \ion{Ca}{ii} lines were also removed, as explained in Paper~I.

The final spectra, containing 957 pixels, were normalised by iteratively fitting a second-order polynomial to the pseudo-continuum, 
asymmetrically clipping off points that lay far from the fitting curve. This first normalisation is not very crucial, because the 
pipeline that derives the atmospheric parameters renormalises the spectra with the aid of synthetic spectra (see Paper~I). 

\begin{figure}
  \resizebox{\hsize}{!}{\includegraphics{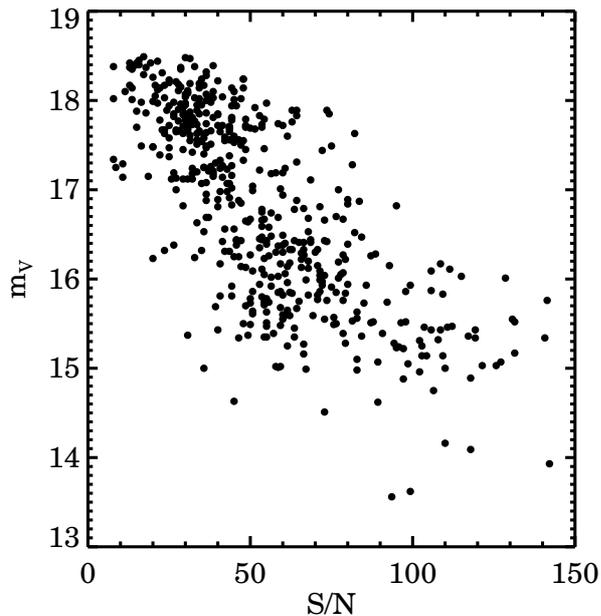}}
  \caption{Apparent magnitude $m_v$ versus measured signal-to-noise ratio (per pixel) for all observed sample (689 stars in total).}
  \label{fig:Mag_vs_SNR}
\end{figure}

Finally, the binary stars or suspected ones, and very low quality spectra (owing to a bad location of the fibre on the target) were 
removed at this point from the rest of the sample, leading to a final sample of 636 stars. The final measured S/N of our spectra varies from 
$\sim 5$ to $\sim 200~\rm{pixel}^{-1}$, with a mean of $\sim 70~\rm{pixel}^{-1}$. 
Nevertheless, the cumulative smoothing effects of spectral interpolation and resampling lead to an over-estimation of the S/N by a factor of $\sim 1.4$ because of the pixel correlation. Hence, the plot of  Fig.~\ref{fig:Mag_vs_SNR} shows the apparent magnitude versus the corrected S/N.

%
%________________________________________________________________

\section{Determination of the stellar atmospheric parameters}
\label{sec:pipeline_atm}
We used the procedure described in Paper~I to derive the atmospheric parameters of the stars and their associated errors. 
To take into account the overestimation of the S/N owing to the pixel resampling, the measured S/N was decreased by a factor 1.4 at each step of the pipeline.   
The obtained parameter values are shown in the online Table~\ref{online}.

We recall that this procedure consists  of using two different algorithms simultaneously, MATISSE \citep{Matisse_MNRAS} and DEGAS 
\citep{Bijaoui_automatic_classification_methods},  to iteratively renormalise the spectra and derive the atmospheric 
parameters of the observed targets. The learning phase for the algorithms is based on  a grid of synthetic spectra covering 
$T_\mathrm{eff}$ from 3000~K to 8000~K, $\log~g$ from 0 to 5 (\mbox{cm s}$^{-2}$) and [M/H] from $-5$~dex to +1.0~dex.
A coupling between the overall metallicity and the $\alpha$ element 
abundances\footnote{The chemical species considered as $\alpha$-elements are \ion{O}{}, \ion{Ne}{}, \ion{Mg}{}, 
\ion{Si}{}, \ion{S}{}, \ion{Ar}{}, \ion{Ca}{} and \ion{Ti}{}.} is assumed according to the commonly observed 
enhancements in metal-poor galactic stars:
\begin{itemize}
\item $[\alpha$/Fe]=0.0 dex for  [M/H] $\geq$0.0 dex
\item $[\alpha$/Fe]=$-0.4 \times $[M/H] dex for $-1\leq$[M/H]$< 0$~dex
\item $[\alpha$/Fe]=+0.4 dex for [M/H] $\leq$ $-$1.0  dex.

\end{itemize}

At S/N$\sim$50$~\rm{pixel}^{-1}$ the pipeline returns typical errors\footnote{defined as the 70\% value of the internal error distribution}
 for stars with $\log~g \geq$3.9 and $-0.5 <$[M/H] $\leq-0.25$~dex,  
of 70~K, 0.12~dex, 0.09~dex for $T_\mathrm{eff}$, $\log~g$ and [M/H], respectively. For more metal-poor stars   
($-1.5 <$[M/H] $\leq-0.5$~dex) the achieved accuracies are 108~K, 0.17~dex and 0.12~dex. Finally, for stars classified as  halo giants 
($T_\mathrm{eff}<$6000~K, $\log~g<$3.5, $-2.5 <$[M/H] $\leq -1.25$~dex) the typical errors are 94~K, 0.28~dex and 0.17~dex (see Paper~I). 

This pipeline was successfully tested  on the observed stellar libraries of $S^4N$ \citep{Prieto_S4N} and CFLIB \citep{CFLIB}, showing no 
particular biases in $T_\mathrm{eff}$ or $\log~g$. Still, a bias of $-0.1$~dex was found for metal-rich dwarfs, which, 
as discussed in Paper~I, we decided not to correct because of the small range of $\log~g$ and [M/H] spanned by the libraries. 
As discussed in Sect.~\ref{subsec:distances}, this possible bias is not expected to introduce any significant bias in the distance estimates
or the derived velocities of the present survey.

%
%________________________________________________________________
 \section{Determination of the stellar distances, kinematics and orbits}
\label{sec:dynamics}
 Typical F, G and K main-sequence stars have $m_\mathrm{v} \sim 15-18$ at distances of 1-5~kpc, and at least until the ESA/Gaia mission, 
stellar distances for targets far from the solar neighbourhood have to be determined spectroscopically or photometrically.
For instance, the atmospheric parameters determined in the previous section can be projected onto a set of theoretical isochrones to derive the absolute 
magnitudes of the stars. Then we can derive the line-of-sight distances using the distance modulus. 

\subsection{Procedure to estimate stellar distances}
\label{subsec:distances}
We generated our own set of isochrones  by  using the \textit{YYmix2}  interpolation code, based on the Yonsei-Yale ($Y^2$) models 
\citep[version~2,][]{Demarque04} combined with the \citet{Lejeune_color_table} colour table. 

We set the youngest age of the isochrones at 2~Gyr, because we expect to observe targets from the old thin disc, the thick disc and 
the halo, and very few stars are expected to be younger than this. Following \citet{Zwitter_distances}, we generated 
isochrones with a constant step of 1~Gyr, up to 14~Gyr.  In addition, we used the full metallicity range of the $Y^2$ models, 
ranging from [Fe/H]=$-$3.0~dex to [Fe/H]=+0.8~dex. The step in metallicity is constant, equal to 0.1~dex, smaller than the typical 
error on the derived metallicities of the observed stars. The adopted values for the $\alpha$-enhancements at different metallicities  
are the same as those used for the grid of synthetic spectra described in the previous section. 
At the end, a set of 494 isochrones were generated.

To obtain the absolute magnitude $M_\mathrm{v}$, we used the method of \citet{Zwitter_distances}. This procedure consists of 
finding the most likely values of the stellar parameters, given the measured atmospheric ones, and the time spent by a star in 
each region of the H--R diagram.
In practice,  we selected the subset of isochrones with [M/H]$\pm \Delta_\mathrm{[M/H]}$,
where $\Delta_\mathrm{[M/H]}$ is the estimated error on the metallicity, for each set of derived $T_\mathrm{eff}$, $\log~g$ and [M/H]. 
Then a Gaussian weight was associated to each point of the selected isochrones, which depends on the measured atmospheric parameters 
and the considered errors (see Eq.~\ref{eqn:weight_isochrones}). This criterion allows the algorithm to select only the points whose 
 values are close to those derived by the pipeline.
 Still, in low S/N spectra, whose parameter errors can be significant, an excessive weight can be appointed to rapid evolutionary phases of the isochrones, where it is unlikely that there are many stars.  To avoid this artefact, we associate to the previous weight, another one, 
 that is proportional to the time spent by a star on each part of the H--R diagram. Following \citet{Zwitter_distances}, this weight was set as $dm$: 
the mass step between two points of the same isochrone.  Hence, the total weight $W$ for each point on the isochrone can be expressed as follows:

\begin{equation}
W=dm \cdot \mathrm{exp}\left(-\sum_i \frac{(\theta_i - \hat \theta_i)^2}{2\Delta^2_{\hat \theta_i}}\right),
\label{eqn:weight_isochrones}
\end{equation}
where $i$ corresponds to $T_\mathrm{eff}$, $\log~g$, [M/H] or (B-V), $\theta_i$  to the parameter values of the points on the isochrones, 
$\hat \theta_i$ to the values derived from observations (from photometry or spectroscopy) and $\Delta_{\hat \theta_i}$ to the associated error of the measurement. 
The absolute magnitude $M_\mathrm{v}$ is then obtained by computing the weighted mean of all points of the subset of isochrones. The corresponding error is computed by considering the standard deviation of the latter. \\

To test the adopted procedure, we made extensive use of the Besan\c{c}on model of the Milky Way \citep{Modele_besancon}. 
This is a semi-empirical model that includes physical constraints and current knowledge of 
the formation and evolution of the Galaxy to return for a given line-of-sight (\textit{los}) the atmospheric 
parameters, the positions and the kinematics of simulated stars in a given range of magnitudes. 
It uses  the isochrones of \citet{Schaller_isochrones} to simulate the stellar populations of the thin disc and the isochrones 
of \citet{isochrones_besancon} to simulate the thick disc and the halo. Thus, small offsets may be expected when comparing the distances 
we derived with the $Y^2$ isochrones and those deduced from this model.

We obtained a simulated catalogue of $\sim 4\cdot10^3$ stars towards ($l \sim 277^{\circ}$, $b \sim 47^{\circ}$) in this way and tested the method 
by computing the \textit{los} distances of the targets in two ways:

\begin{enumerate}
\item  The Besan\c{c}on atmospheric parameter values were taken and associated with errors that the pipeline of Paper~I would 
produce for the corresponding stellar type and metallicity at different S/N. 
In that  way we checked the impact of the weight $W$ of Eq.~\ref{eqn:weight_isochrones} on the distance estimation.
We represented in the first row of Table~\ref{tab:q70_besancon} the values at 70\% of the error distribution that we found.
The results are very encouraging because at S/N $\sim20~\rm{pixel}^{-1}$ the dispersion of the recovered distances with respect to the theoretical 
ones is only 27\%. Nevertheless, a bias was found for  stars farther than $\sim$8~kpc,  
even at high S/N. For instance, at S/N$\sim50~\rm{pixel}^{-1}$ the distances for the farthest stars are overestimated by $\sim$18\%, 
and this offset increases with decreasing S/N (up to 30\% at S/N$\sim10~\rm{pixel}^{-1}$). 
Indeed, these distant stars are found to be metal-poor giants (halo stars), for which the errors in the 
atmospheric parameters are expected to be larger (e.g.: at S/N$\sim20~\rm{pixel}^{-1}$, the uncertainties are 188~K, 0.57~dex and 
0.23~dex for $T_\mathrm{eff}$, $\log~g$ and [M/H], respectively). 
These uncertainties will allow a high number of points on the isochrones to have roughly the same weight $W$ and to span
a relatively wide range of magnitudes, because the late-type giant stars have a sensitive dependance of absolute 
magnitude with astrophysical parameters. This effect will result in a mean $M_\mathrm{v}$ that will be lower, designating the stars 
to a farther distance than they should be. This problem is well known and has already been described in detail in \citet{Breddels_distances} 
and \citet{Zwitter_distances}.
As we will see in Sect.~\ref{subsec:halo_selection}, it will concern less than 4\% of our observed sample. 

A disagreement of $\sim$13\% towards smaller distances for the stars 
closer than 1~kpc was found. This difference is caused by the different sets of isochrones involved, and not by the method itself. 
Indeed, we verified that the absolute magnitudes for main-sequence stars at a given metallicity were always higher in the case of 
\citet{Schaller_isochrones} compared to the $Y^2$ ones. The difference disappeared for the dwarfs generated from the isochrones of 
\citet{isochrones_besancon}.

\item The parameters derived from the spectra of the pseudo-stars were taken. For that purpose, we computed degraded (S/N$\sim$100, 50, 20, 10$~\rm{pixel}^{-1}$) 
synthetic spectra with the atmospheric parameters of the simulated targets and used the automated pipeline to derive their  $T_\mathrm{eff}$, $\log~g$ and [M/H]. 
We found that the dispersion on the final distance estimation was slightly increased, and that no additional biases were introduced. 
Indeed, the projection on the isochrones fixes some possible misclassification of the automated spectral classification pipeline. 
For instance, it solves the problem of the thickening of the cool part of the main-sequence, as presented in Paper~I.
Therefore, the estimations are quite similar to those obtained from the Besan\c{c}on values, as seen in the last row of Table~\ref{tab:q70_besancon}.

\end{enumerate}

Finally, we tested the effect of a possible metallicity bias on the distance determinations. We modified the Besan\c{c}on metallicities 
by $-0.1$~dex to match the suspected bias that the pipeline of Paper~I exhibited. The errors on the recovered distances were only 
increased by 2\% compared to those presented in Table~\ref{tab:q70_besancon}. 
We therefore conclude that our distance estimates are not sensitive to the possible small bias in metallicity that could be present in our data.

\addtocounter {table} {2}
\begin{table} 
\centering
\caption{Values at 70\% of the error distribution for the recovered distances of the pseudo-stars of the Besan\c{c}on model. }
\begin{tabular}{lcccc}
\hline
\hline
S/N ($~\rm{pixel}^{-1}$)  & 100 &  50 &  20 &  10 \\\hline
Relative errors (1)     & 17\% & 20\%  & 27\%   & 31\% \\
Relative errors (2)     & 19\% & 23\%  & 35\%   & 52\% \\ \hline
\end{tabular}
\label{tab:q70_besancon}
\tablefoot{Recovered distances (1) using the Besan\c{c}on atmospheric parameters and (2) using the parameters derived 
from interpolated synthetic spectra at the Besan\c{c}on parameters.}
\end{table}

\subsection{Distances of the observed stars}
\begin{figure}
  \resizebox{\hsize}{!}{\includegraphics{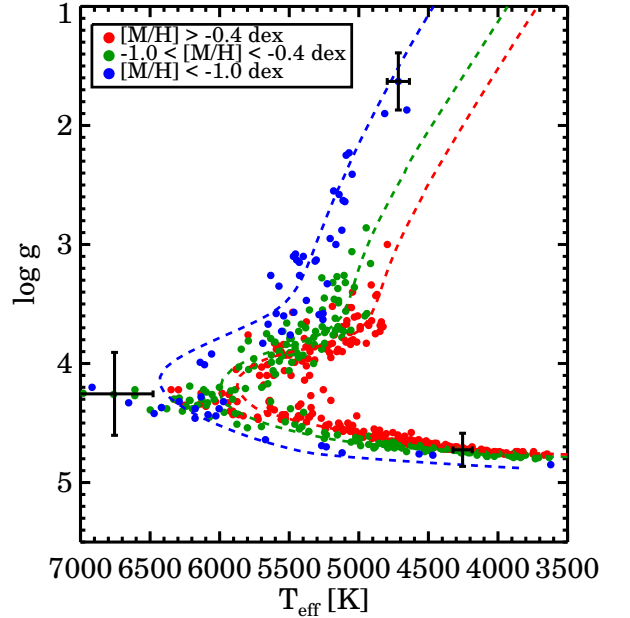}}
  \caption{Atmospheric stellar parameters for the present sample of galactic field stars derived by the pipeline 
    of Paper~I and  projected onto the $Y^2$ isochrones. Three isochrones for [M/H]=--1.5~dex and 13~Gyr (blue), 
    [M/H]=--0.5~dex and 10~Gyr (green) and [M/H]=0~dex and 7~Gyr (red) are represented. The metallicity values 
    are those derived by the automated pipeline. Typical error bars are represented for different regions of the H--R 
    diagram. }
  \label{fig:HR_diagram}
\end{figure}

\begin{figure}
  \resizebox{\hsize}{!}{\includegraphics{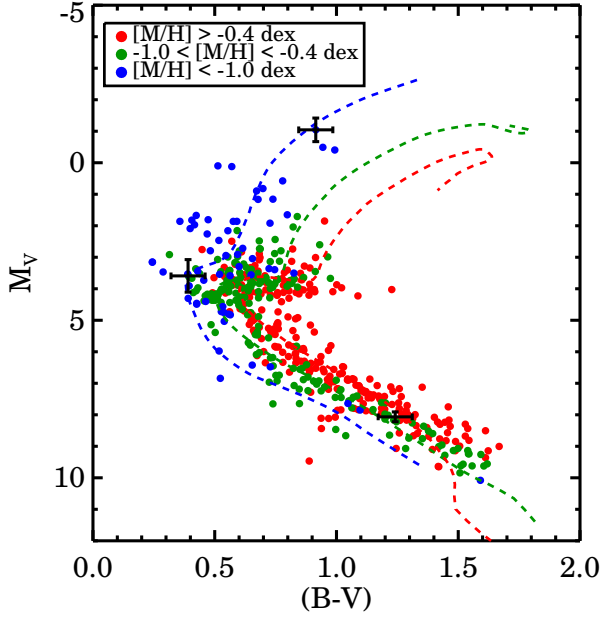}}
  \caption{ Derived absolute magnitude $M_V$ versus (B-V) colour for the present sample of galactic field stars and for the atmospheric parameters  derived by the pipeline  of Paper~I and  projected on the $Y^2$ isochrones. The colour codes and the isochrones are the same as in Fig.~\ref{fig:HR_diagram}. Typical error bars are represented for different regions of the H--R diagram. }
  \label{fig:CMD_diagram}
\end{figure}

\begin{figure}
  \resizebox{\hsize}{!}{\includegraphics{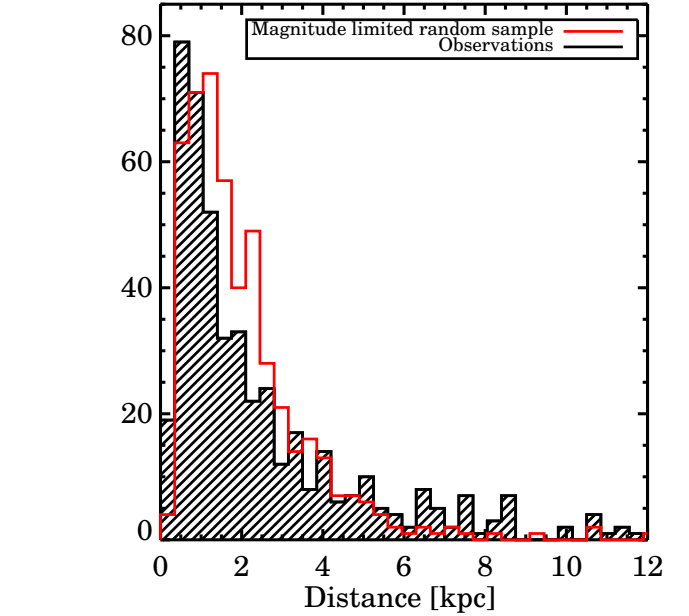}}
  \caption{ In black we denote the distribution of the line-of-sight distances for the subset of the observed stellar sample that meet the quality 
    criteria described in Sect.~\ref{subsec:star_selection}. Overplotted in red is the distribution of a randomly selected magnitude-limited ($m_V \leq 18.5)$ sample of stars based on the Besan\c{c}on catalogue. }
  \label{fig:los_dist_histogram}
\end{figure}

 The observed magnitudes and colours of \cite{Ojha_photometry} were deredenned assuming a mean $E(B-V) \sim 0.04$~mag and 
$A_\mathrm{v} \sim 0.1$~mag \citep{Extinction_map}. 
Figures~\ref{fig:HR_diagram}, \ref{fig:CMD_diagram} and \ref{fig:los_dist_histogram} show the atmospheric parameters projected onto the isochrones, their absolute magnitude $M_V$, and the 
histogram of the distances that we obtained using the method described in the previous section.
 These figures concern the final sample, as selected in Sect.~\ref{subsec:star_selection}.
 The value of $A_\mathrm{v}$ should be lower for the closest stars than the adopted one because $A_\mathrm{V}$ 
represents the total extinction in the {\it los}. However, the nearest targets are at  $Z\sim$130~pc ($D\sim175$~pc), higher than 
most of the dust column, and consequently a single value of $A_\mathrm{v}$ is a good approximation. In addition,  
a maximum error of 0.1~mag in $A_\mathrm{v}$ will be equivalent to an underestimation of less than 4\% in the distances, 
which can be neglected.

The availability of the distances ($D$) then allows us to estimate their galactic-centred cartesian coordinates\footnote{We 
have adopted a right-handed reference frame, with the X axis pointed towards the galactic centre.} ($X_{GC},Y_{GC},Z_{GC}$):

\begin{eqnarray}
X_{GC}&=&D \cos(b) \cos(l) - X_\odot  \\
Y_{GC}&=&D \cos(b) \sin(l)\\
Z_{GC}&=&D\sin(b),
\end{eqnarray}
where ($X_\odot, Y_\odot, Z_\odot$) = (8, 0, 0)~kpc \citep{Ro_distance_Reid} and $l$, $b$ are the galactic coordinates.\\
Table~\ref{online2}, available online, presents the derived values and their associated errors (computed analytically, considering no errors on $l$ and $b$).
They will be used in Sect.~\ref{sec:results} to assign a star to a particular galactic component (thin disc, thick disc or halo).

\subsection{Kinematic properties and orbital parameters}
\label{subsec:kinematics}
 Proper motions combined with distance estimates and radial velocities provide the information required to calculate the 
full space motions of any star in the Galaxy. For our sample the radial 
velocities were derived from the observed spectra (see Sect.~\ref{sec:observations}), whereas magnitudes, colours and proper 
motions were taken from \cite{Ojha_photometry}. 
The associated space-velocity components in the galactic cardinal directions, U (towards the galactic centre), 
V (in the direction of the galactic rotation) and W (towards the north galactic pole) were computed for all the stars using the 
following equations:

\begin{eqnarray}
U &=& V_\mathrm{rad} \cos(b) \cos(l) - D \mu_b \cos(l) \sin(b)  - D  \mu_l \sin(l) \\
V &=& V_\mathrm{rad} \cos(b) \sin(l) - D \mu_b \sin(l) \sin(b)  + D  \mu_l \cos(l) \\
W &=& V_\mathrm{rad} \sin(b) + D \mu_b \cos(b),
\end{eqnarray}
where $\mu_l$ and $\mu_b$ are the proper motions on the sky\footnote{One measures $\mu_l=\mu_l^*\cos(b)$, with $\mu_l^*$ the true proper motion in the $l$ direction}. 
We also computed the galactocentric velocities in a cylindrical reference frame. In that case, the velocity components are $V_R$, $V_\phi$, and $V_Z$, with $V_Z=W$. They are computed as follows:
\begin{eqnarray}
V_R &=& \frac{1}{R} \left[ X_{GC} \cdot (U+U_\odot) + Y_{GC} \cdot (V+V_{rot}+V_\odot) \right] \label{eqn:vrho}\\
V_\phi &=& \frac{1}{R} \left[ X_{GC} \cdot (V+V_{rot}+V_\odot) - Y_{GC} \cdot (U+U_\odot) \right] \label{eqn:vphi},
\end{eqnarray}
where $R=\sqrt{X_{GC}^2+Y_{GC}^2}$ is the planar radial coordinate with respect to the galactic centre, $(U_{LSR}^{\odot},V_{LSR}^{\odot},W_{LSR}^{\odot})=(10.00,5.25,7.17)$~\mbox{km s}$^{-1}$  is the solar motion decomposed into its cardinal directions  relative to the Local Standard of Rest \citep[LSR,][]{Solar_motions}, and $V_{rot}=220$ \mbox{km s}$^{-1}$ is the amplitude of the galactic rotation towards $l=90^{\circ}$ and $b=0^{\circ}$  \citep[IAU 1985 convention; see][]{Galactic_constants}. In this reference frame a retrograde rotation is indicated by $V_\phi>0$~\mbox{km s}$^{-1}$.

The errors on these kinematic data are estimated as follows:
for each star, we performed $5\cdot10^3$ Monte-Carlo realisations for $D$, $\mu_l$, $\mu_b$ and $V_\mathrm{rad}$, assuming Gaussian distributions 
around their adopted values, with a dispersion according to their estimated errors. For every realisation, 6d phase-space 
parameters and angular momenta were computed, taking as a final value the mean of all realisations, and as an error the standard 
deviation (see Tables~\ref{online2} and~\ref{online3}).  The number of realisations was determined  to achieve a stable result for each measurement. 
Later on, as described in Sect.~\ref{subsec:components}, each of these Monte-Carlo realisations will be used to assign 
for each star a probability of belonging to one of the galactic components.
 
 The median estimated errors on the galactic rotational velocities, $V_\phi$, for metal-rich ([M/H]$>-0.4$~dex), intermediate 
metallicity ($-1<\mathrm{[M/H]}<-0.4$~dex) 
and low-metallicity ([M/H]$<-1$~dex) stars are of the order of  10, 18 and 48~\mbox{km s}$^{-1}$, respectively. 
For $V_R$ we reach accuracies of  12, 23 and 49~\mbox{km s}$^{-1}$, and for $V_Z$ we obtain 9, 16 and 41~\mbox{km s}$^{-1}$, respectively.\\

Finally, from the present positions and space-motion vectors of the stars one can integrate their orbits up to several galactic revolutions and hence estimate their orbital parameters. 
Stellar orbital eccentricities\footnote{We define the eccentricity as $\epsilon=(r_{ap}-r_{pe})/(r_{ap}+r_{pe})$ where $r_{ap}$ and $r_{pc}$ correspond to the apocentric and pericentric distances of the last orbit of each star.}, 
pericentric and apocentric distances were computed. For that purpose we considered a three-component Galaxy (dark halo, disc, bulge). The Milky Way potential was fixed, and modelled with a Miyamoto-Nagai disc, a Hernquist bulge and a  logarithmic dark matter halo \citep[see][and references therein for more details]{Helmi04}. 

\begin{figure}
  \resizebox{\hsize}{!}{\includegraphics{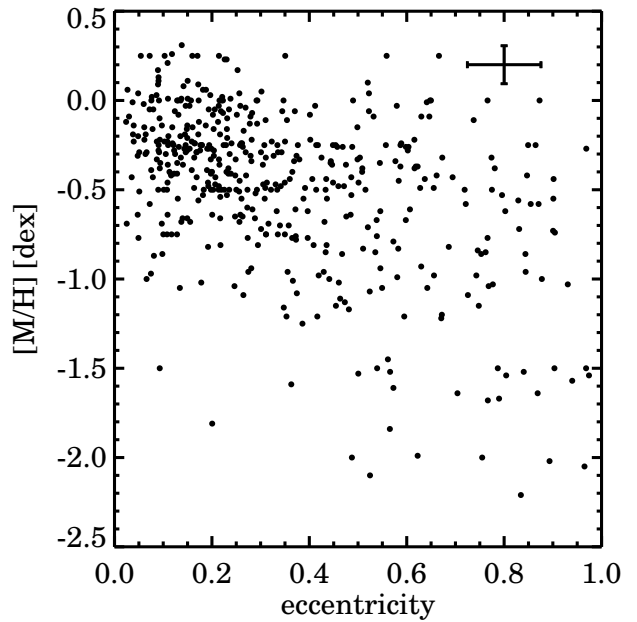}}
  \caption{Metallicity versus eccentricity for the  452 stars with full 6d phase-space coordinates. Typical errors are shown in the upper right corner.}
  \label{fig:FeH_vs_eccen}
\end{figure}

 Errors on these parameters were propagated from the previously found positions and velocities. Typical errors on eccentricities are $\sim$22\%. Results can be seen in Table~\ref{online3} and in Fig.~\ref{fig:FeH_vs_eccen}, where $\epsilon$ has been plotted versus [M/H]. 
As we will discuss in the next section, these eccentricities can be used to identify the galactic components and distinguish the origins of the thick disc stars \citep{Sales_2009}. 
%Following \citet{Sales_2009}, stars having  low eccentricity ($\epsilon < 0.1$) can be assimilated to the thin disc, whereas stars having $\epsilon > 0.7$ belong preferentially to the halo. Stars with intermediate eccentricities are thought to probe the thick disc. A discussion concerning the shape of the eccentricity distribution will be done in Sect.~\ref{subsec:components}. 

%
%________________________________________________________________________
\section{Star selection}
\label{subsec:star_selection}

\begin{figure}
  \begin{center}
    \begin{tabular}{c}
      \includegraphics[width=7.4cm,height=7.4cm]{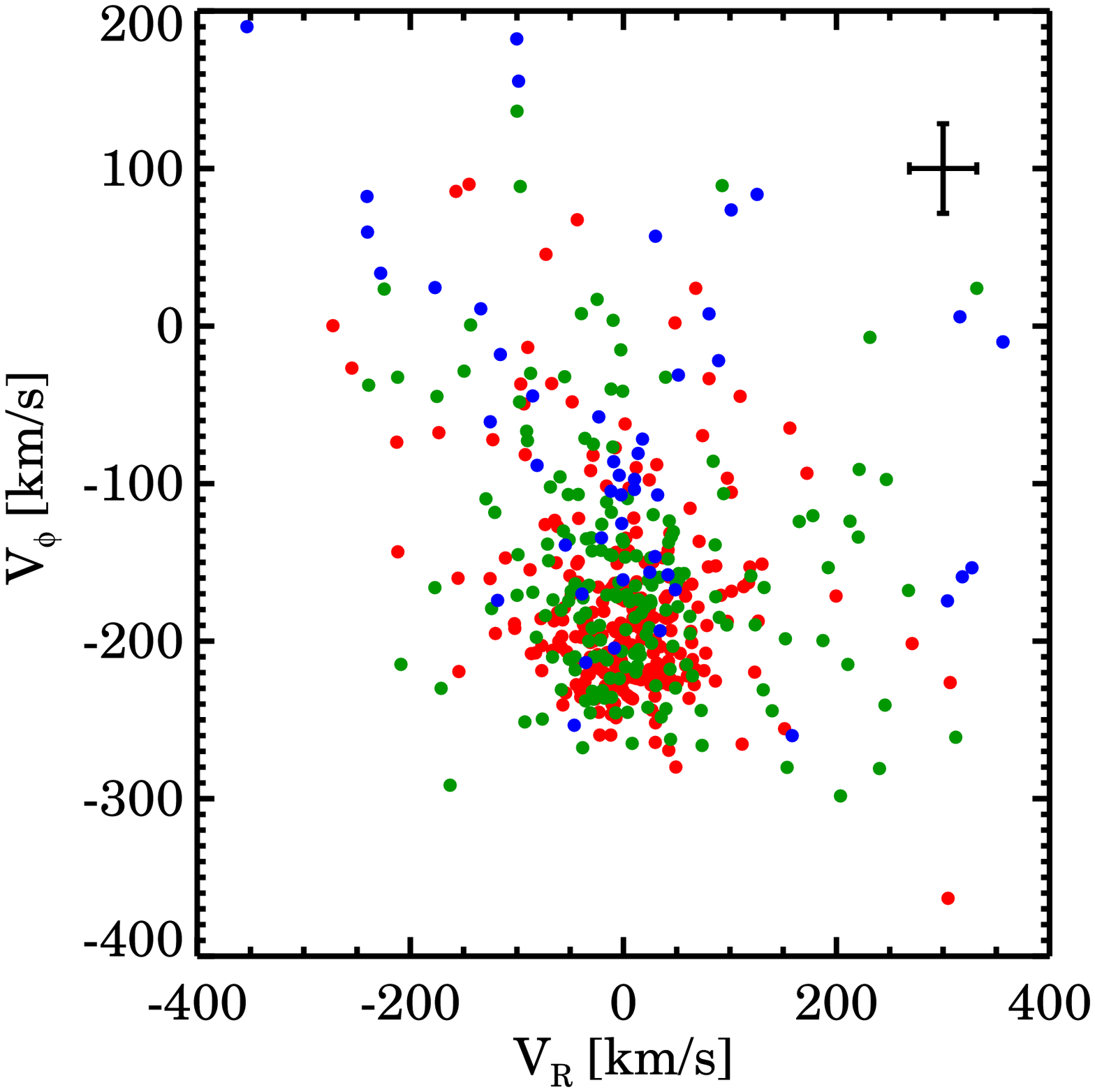} \\
      \includegraphics[width=7.4cm,height=7.4cm]{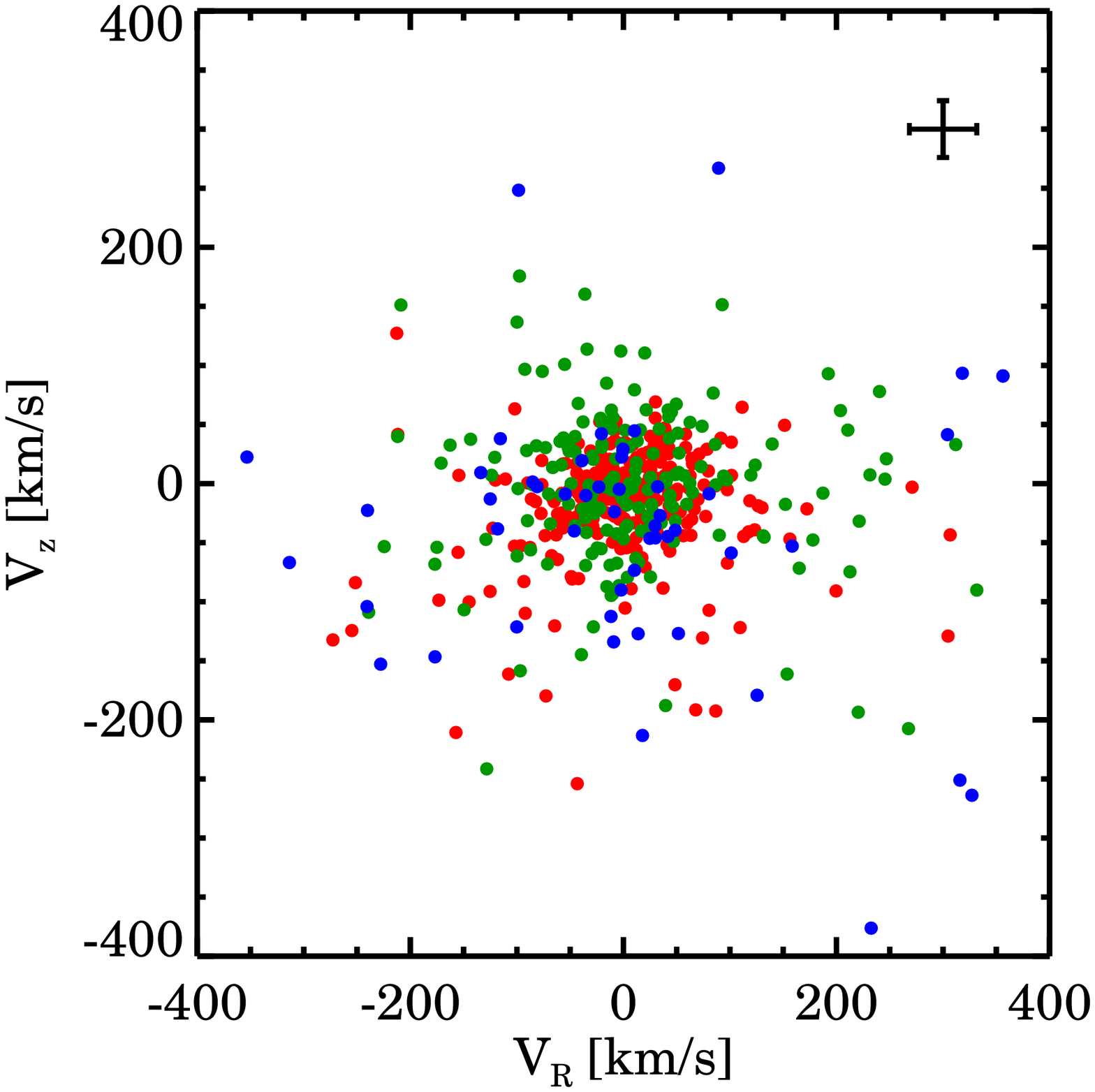} \\
      \includegraphics[width=7.4cm,height=7.4cm]{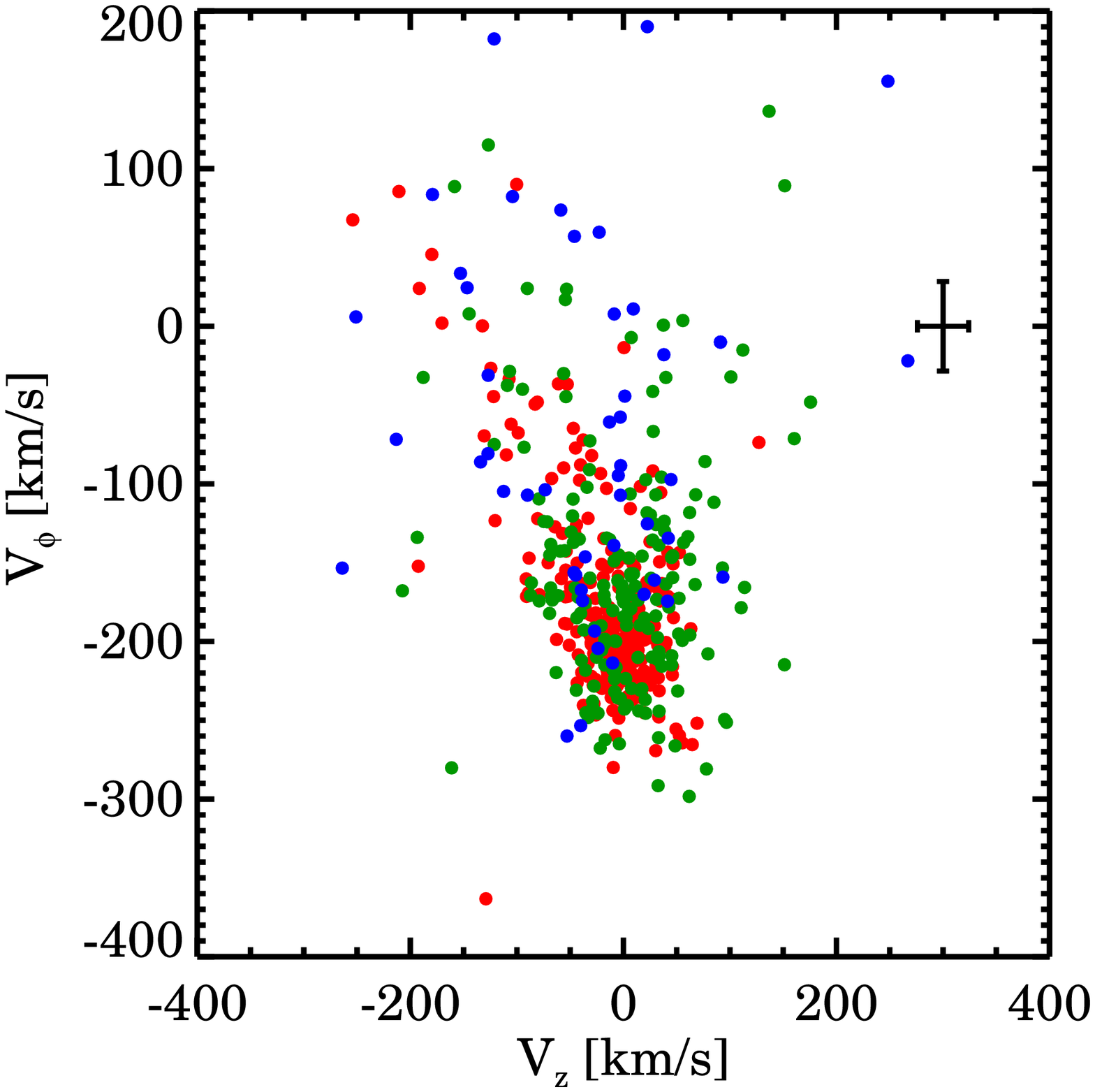} 
    \end{tabular}
\caption{Galactocentric velocities computed for the observed sample as defined in Sect.~\ref{subsec:star_selection}.  Typical errors are represented in the upper right corners. The colour code according to the metallicity is the same as in Fig.\ref{fig:HR_diagram}.}
\label{fig:plots_FLAMES}
  \end{center}
\end{figure}

\begin{figure}
  \begin{center}
    \begin{tabular}{c}
      \includegraphics[width=7.4cm,height=7.4cm]{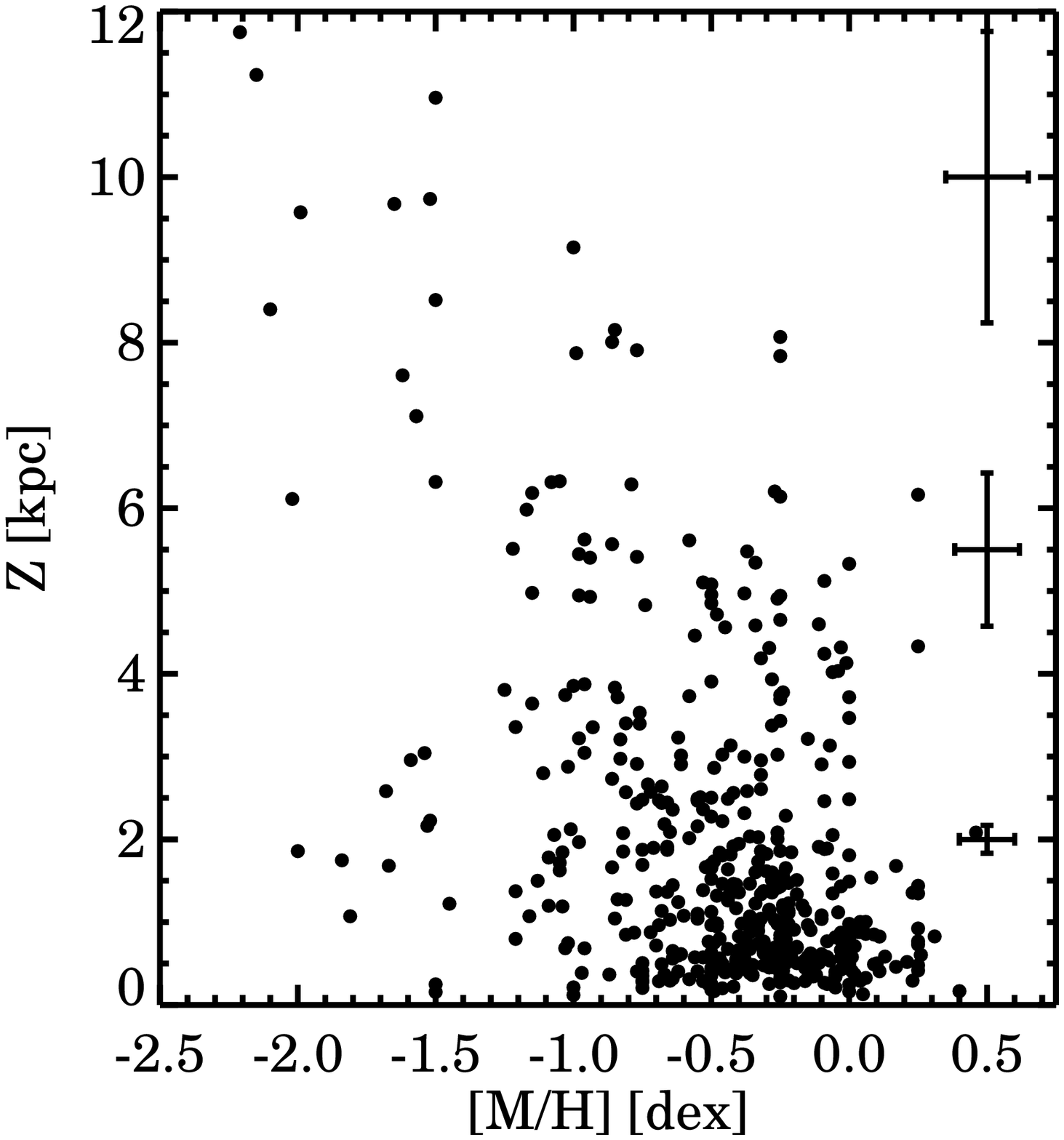} \\
      \includegraphics[width=7.4cm,height=7.4cm]{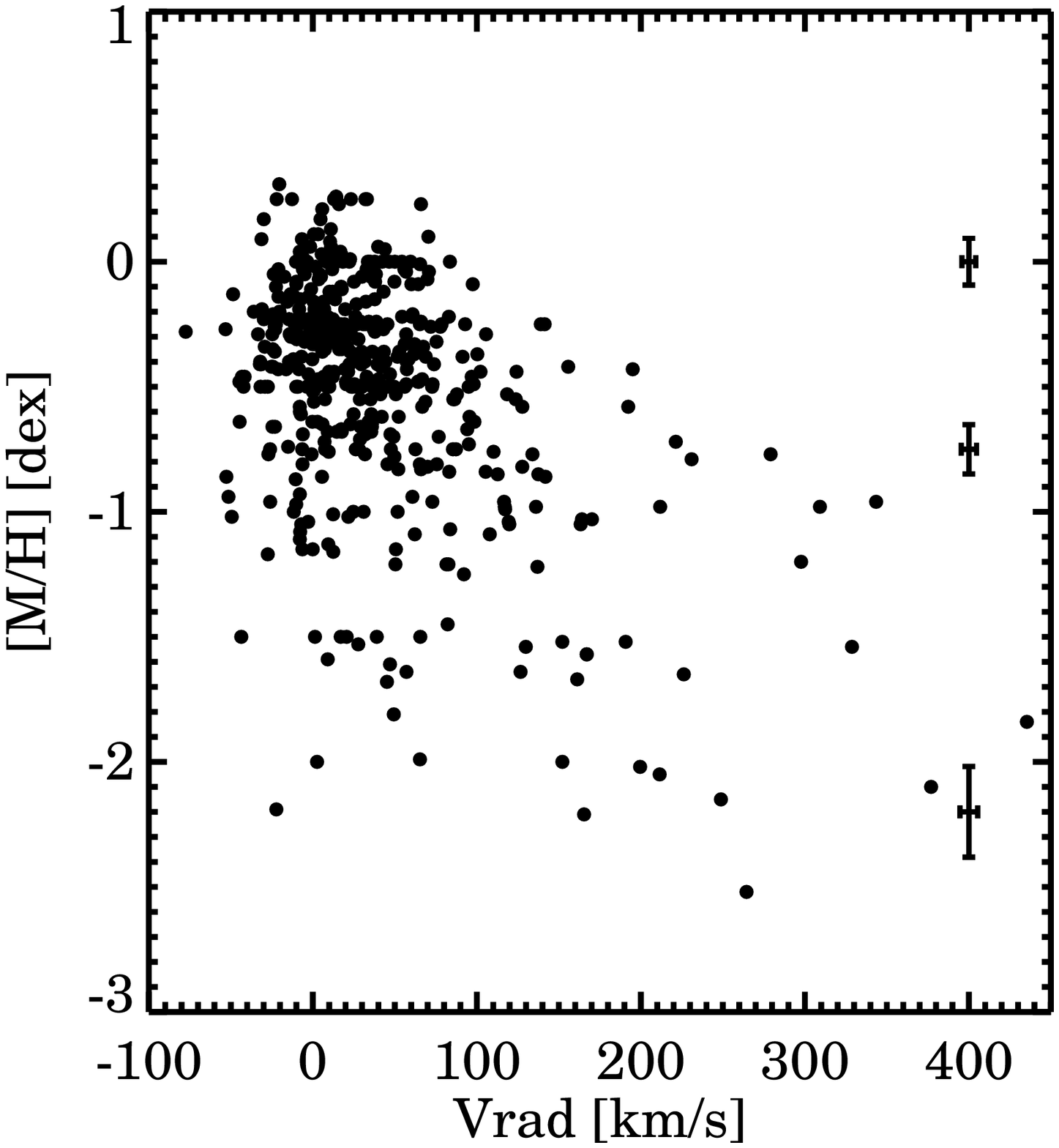} \\
      \includegraphics[width=7.4cm,height=7.4cm]{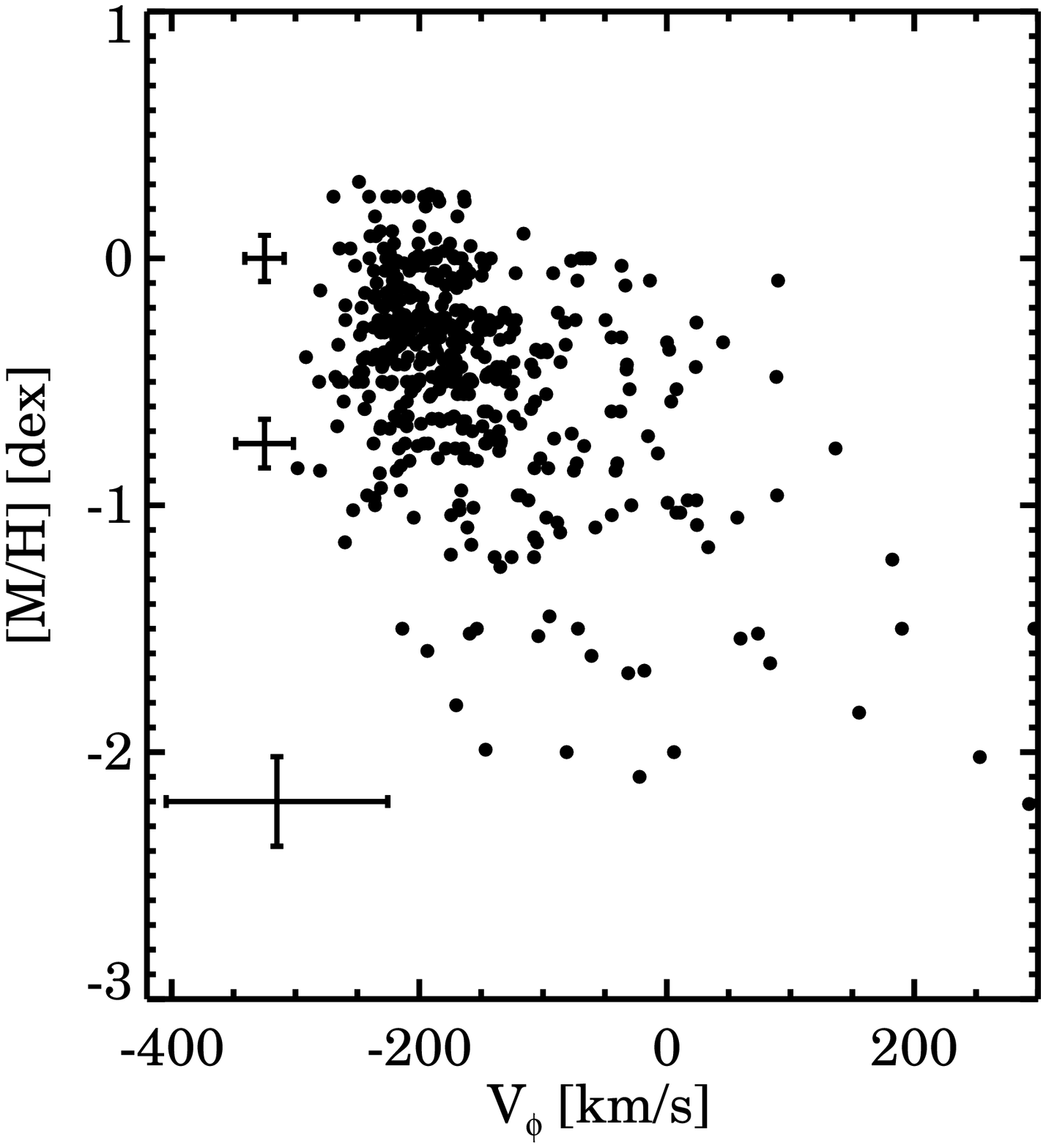} 
    \end{tabular}
    \caption{Metallicity versus distance above the plane (top), versus heliocentric radial velocity ($V_\mathrm{rad}$, middle plot) and versus rotational velocity ($V_\phi$, lower plot), for the considered sample. Mean estimated errors are represented for three different heights or metallicity ranges.}
  \label{fig:FeH_vs_velocities}
  \end{center}
\end{figure}

We need the cleanest possible sample to confidently describe the properties of our observed stars. 
The final catalogue contains  479 stars, 452 of which have full 6d phase-space coordinates. Below we explain 
how they have been selected.

 As already mentioned, from the initial 689 targets 53 stars that were identified  as binaries or had very low quality 
 spectra were removed. 
 In addition, we removed the targets for which the parameters projected onto the isochrones clearly disagreed with the results of the pipeline, according to the following criterion:
\begin{equation}
\frac{\chi^2_\mathrm{isochrone}-\chi^2_\mathrm{pipeline}}{\chi^2_\mathrm{pipeline}} > 0.25 ,
\end{equation}
where $\chi^2_\mathrm{pipeline}$ and $\chi^2_\mathrm{isochrone}$ correspond to the differences between the observed spectrum 
and the synthetic templates (resulting from the pipeline parameters or those projected onto the isochrones, respectively). 
 The above condition removed  24 stars, mainly giants.
In addition,  11 stars with a projection onto an isochrone of 2~Gyr were removed from our survey because these young stellar populations
are expected neither in the thin disc (at those distance above the plane), the thick disc or the halo. 
Removing these targets is justified, because the stars that are found to be so young are clearly misclassified by the 
automatic pipeline (because of degraded spectral lines, bad normalisation..., or because of the $\log~g$-$T_\mathrm{eff}$ degeneracy between the 
hot dwarfs and the giants, see Paper~I). In addition, this criterion also removes possible stars on the blue horizontal branch (BHB), 
because  the $Y^2$ isochrones  closest to the BHB will be the youngest ones.   
Finally, we removed 11 stars for which the spectra had an estimated error in the $V_\mathrm{rad} > 10~\mbox{km s}^{-1}$,  43 stars 
with an estimated error on $D>50\%$, and all remaining stars with S/N$< 20~\rm{pixel}^{-1}$. 
Indeed, as shown in Paper~I, the errors in the atmospheric parameters, and hence on the distances are 
non negligible in that case. As a matter of fact, with a selection for S/N$\geq$20$~\rm{pixel}^{-1}$, errors less than 190~K, 0.3~dex and 0.2~dex are expected 
for thick disc stars for $T_\mathrm{eff}$, $\log~g$ and [M/H], respectively, leading to errors smaller than 35\% on the distances.

Figures~\ref{fig:HR_diagram} and  \ref{fig:los_dist_histogram} show the H--R diagram and the {\it los} distance distribution for 
the final catalogue, which contains  479 stars. For the  452 stars for which the proper motions were available, Fig.~\ref{fig:plots_FLAMES} 
illustrates their  $V_R$, $V_\phi$, and $V_Z$ velocity components. 
 The targets mainly span {\it los} distances from $\sim$175~pc up to $\sim$10~kpc (corresponding to a distance above the plane $Z\sim 7.5$~kpc), 
with only  11 stars reaching up to $D\sim$32~kpc ($Z\sim24$~kpc).
As described in Sect.~\ref{subsec:distances}, these very distant stars are thought to have strongly overestimated distances (up to 30\%), and hence one should be careful concerning the conclusions obtained for them. 
In addition, the excess of stars seen in the $V_Z$ versus $V_R$ plot of Fig.~\ref{fig:plots_FLAMES} towards negative $V_Z$ and $V_R$ is caused by small number statistics rather than a stellar stream.

\begin{figure}
  \resizebox{\hsize}{!}{\includegraphics{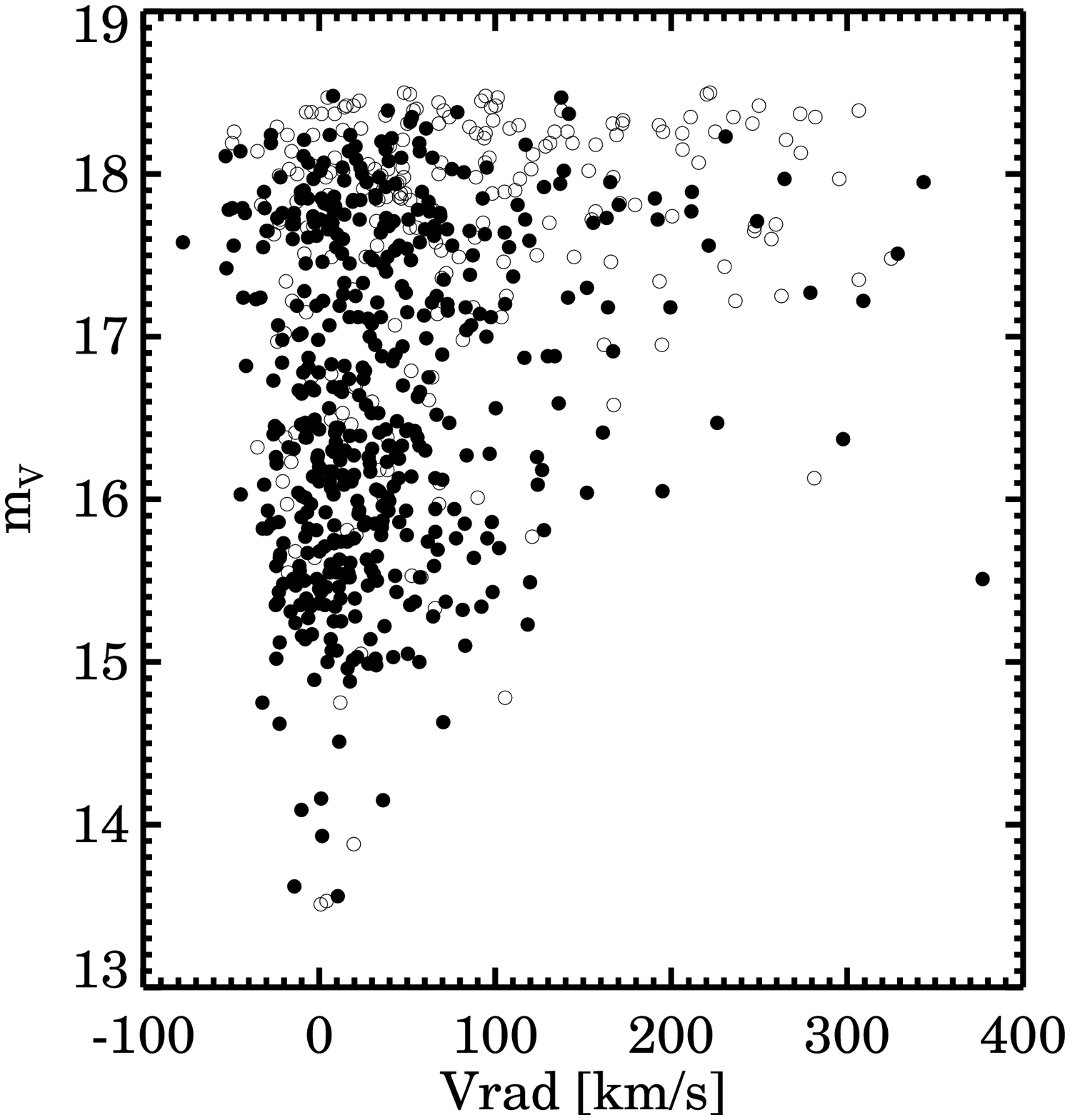}}
  \caption{Apparent magnitude versus derived heliocentric radial velocity ($V_\mathrm{rad}$) of the entire sample. Empty circles represent the rejected targets, whereas the filled ones represent the selected  stars for the final catalogue (see Sect.~\ref{subsec:star_selection} for more details).   Most of the faint targets with $V_\mathrm{rad} \ge 180~\mbox{km s}^{-1}$ (typically, the halo stars) were removed.}
  \label{fig:mv_vs_vrad}
\end{figure}

The metallicity clearly decreases with increasing $Z$ (Fig.~\ref{fig:FeH_vs_velocities}), as expected, because the proportion 
of thin disc, thick disc and halo stars changes.
Nevertheless, we can notice in Fig.~\ref{fig:FeH_vs_velocities} a lack of metal-poor stars ([M/H]$<-$1.5~dex) at heights greater 
than $\sim$3~kpc from the plane. 
One could argue that these missing stars are either misclassified, unobserved, or removed because of one of the quality 
criteria cited previously. The analysis of the  magnitudes and the $V_\mathrm{rad}$ of the removed stars (empty circles in Fig.~\ref{fig:mv_vs_vrad}) 
 shows that these targets are mainly faint stars ($m_v \ge 18$), with typical velocity values corresponding to the halo 
($V_\mathrm{rad} \ge 180~\mbox{km s}^{-1}$). This is  strong evidence suggesting that some metal-poor stars have been observed, 
but were removed afterwards from the final catalogue.
Indeed, the poor accuracies  achieved for this type of stars (spectral signatures lost in the noise, see Paper~I) lead to their 
removal when selecting a clean sample for the present analysis.

Finally, the $V_\mathrm{rad}$ distribution of the targets extends up to $V_\mathrm{rad}=+400~\mbox{km s}^{-1}$ (see Fig.~\ref{fig:FeH_vs_velocities}). At these galactic latitudes ($b\sim47^{\circ}$)  $V_\mathrm{rad}\gtrsim300~\mbox{km s}^{-1}$ corresponds to halo stars with retrograde orbits. The low metallicities of these stars combined with the derived  $V_\phi$-velocity component confirm this.
These stars will be discussed in more detail in Sect.~\ref{subsec:halo_selection}.

%
%________________________________________________________________________
\section{Characterisation of the observed stellar sample}
\label{sec:results}

To compare our observations with galactic models, we created a catalogue of pseudo-stars with the same properties as ours. 
For that purpose we obtained a complete (up to $m_V$=19, $m_B$=20)  catalogue of $8\cdot10^4$ simulated stars towards our {\it los} from the 
website of the Besan\c{c}on galactic model\footnote{\url{http://model.obs-besancon.fr}} \citep{Modele_besancon}. 
The latter supposes a thick disc formed  through one or successive merger processes, resulting in a unique age population (11~Gyr) with a scale height of 800~pc, a local density of 6.2\%, and no vertical gradients in metallicity or rotational velocity.

To run the model, we considered a mean diffuse absorption of 0.7~mag~kpc$^{-1}$. We applied parabolic photometric errors as a function of the magnitude, and a mean error on the proper motions of 2~mas~year$^{-1}$, according to the values given by \citet{Ojha_photometry}. The input errors for the $V_\mathrm{rad}$ were those derived in Sect.~\ref{sec:observations}.  We selected stars with a Monte-Carlo routine to obtain the same $m_V$ and (B-V) distributions as in the FLAMES survey and the same ratio between giant and main-sequence stars (resulting from our final catalogue). This strict selection  finally kept $\sim 4 \cdot 10^3$ simulated Besan\c{c}on stars (called raw Besan\c{c}on catalogue hereafter).
Below we will first use this mock catalogue to help us in interpreting the vertical properties of the observed data (Sect.~\ref{subsec:comparaison_besancon}). Then, according to the hints given by the vertical study, the stars belonging to each galactic component will be selected (Sect.~\ref{subsec:components}), which will lead to a characterisation of the thin disc, thick disc and halo.

We recall that the model returns the cartesian velocities U, V, W, and  to obtain the values $V_R$, $V_\phi$ in the cylindrical frame, one must use Eqs.~\ref{eqn:vrho} and \ref{eqn:vphi}. Both of them suppose a solar velocity and a galactocentric distance for the Sun. The values adopted by the Besan\c{c}on model are not the commonly admitted ones ($R_\odot=8.5$~kpc, $U_\odot=10.3$~\mbox{km s}$^{-1}$, $V_\odot=6.3$~\mbox{km s}$^{-1}$, $W_\odot=5.9$~\mbox{km s}$^{-1}$), and for that reason, we preferred to compare only the cartesian velocities and not the best-suited cylindrical frame.

%
%______________________________________________________________________
\subsection{Study according to the distance above the galactic plane}
\label{subsec:comparaison_besancon}

\begin{figure*}
  \begin{center}
    \begin{tabular}{cc}
      \includegraphics[width=5.6cm,height=5.6cm]{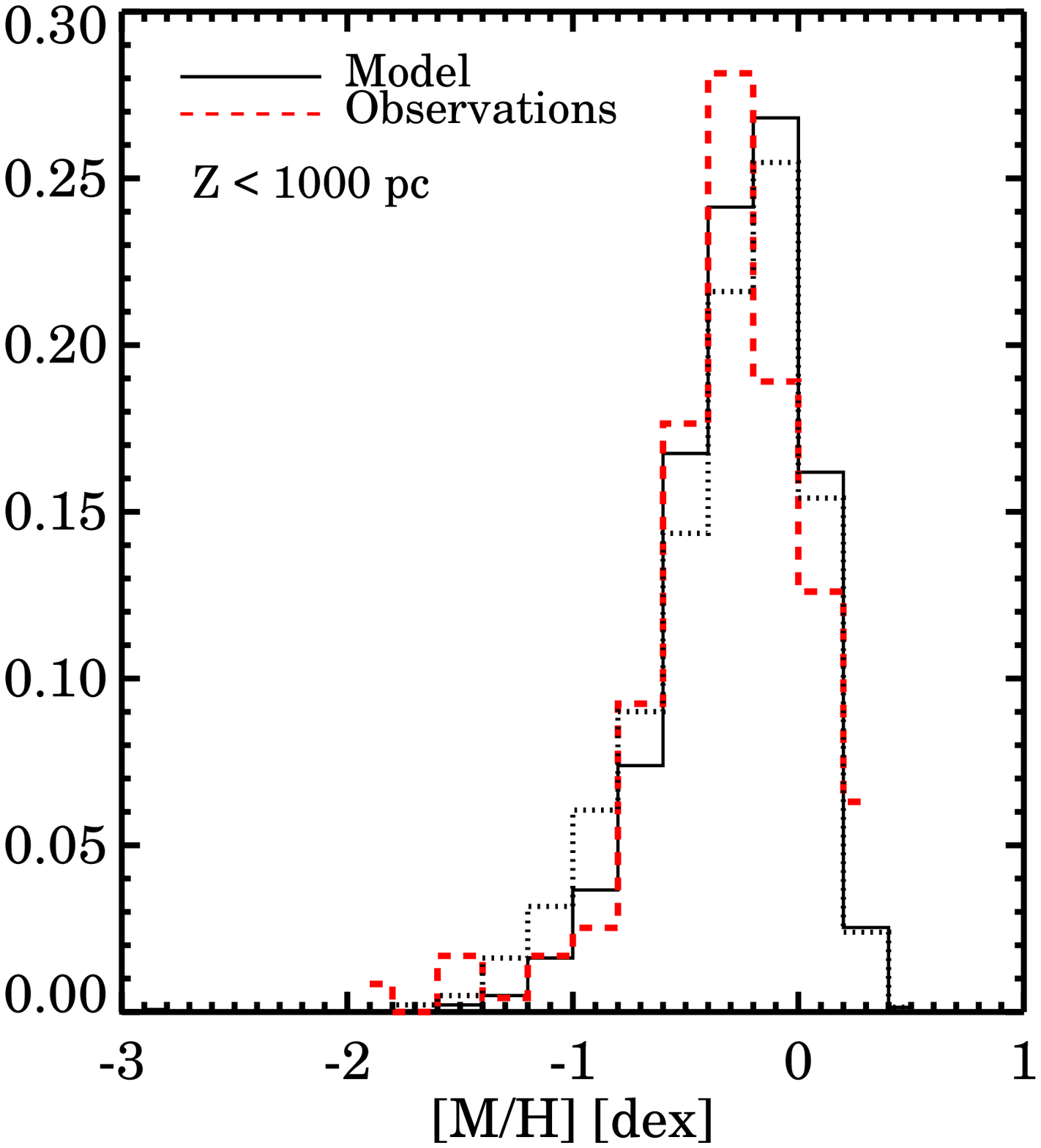} & \includegraphics[width=5.6cm,height=5.6cm]{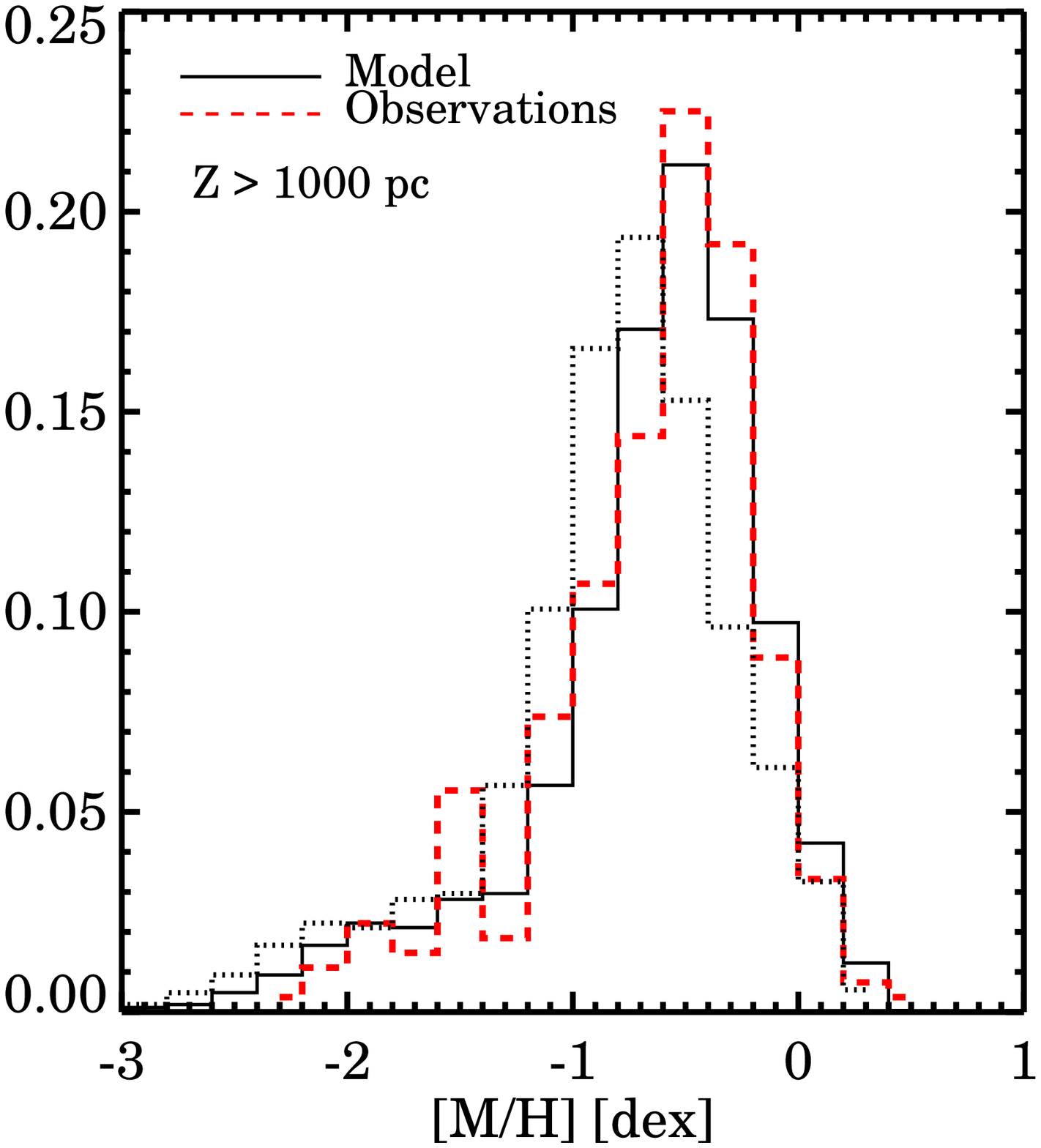}  \\   
      \includegraphics[width=5.6cm,height=5.6cm]{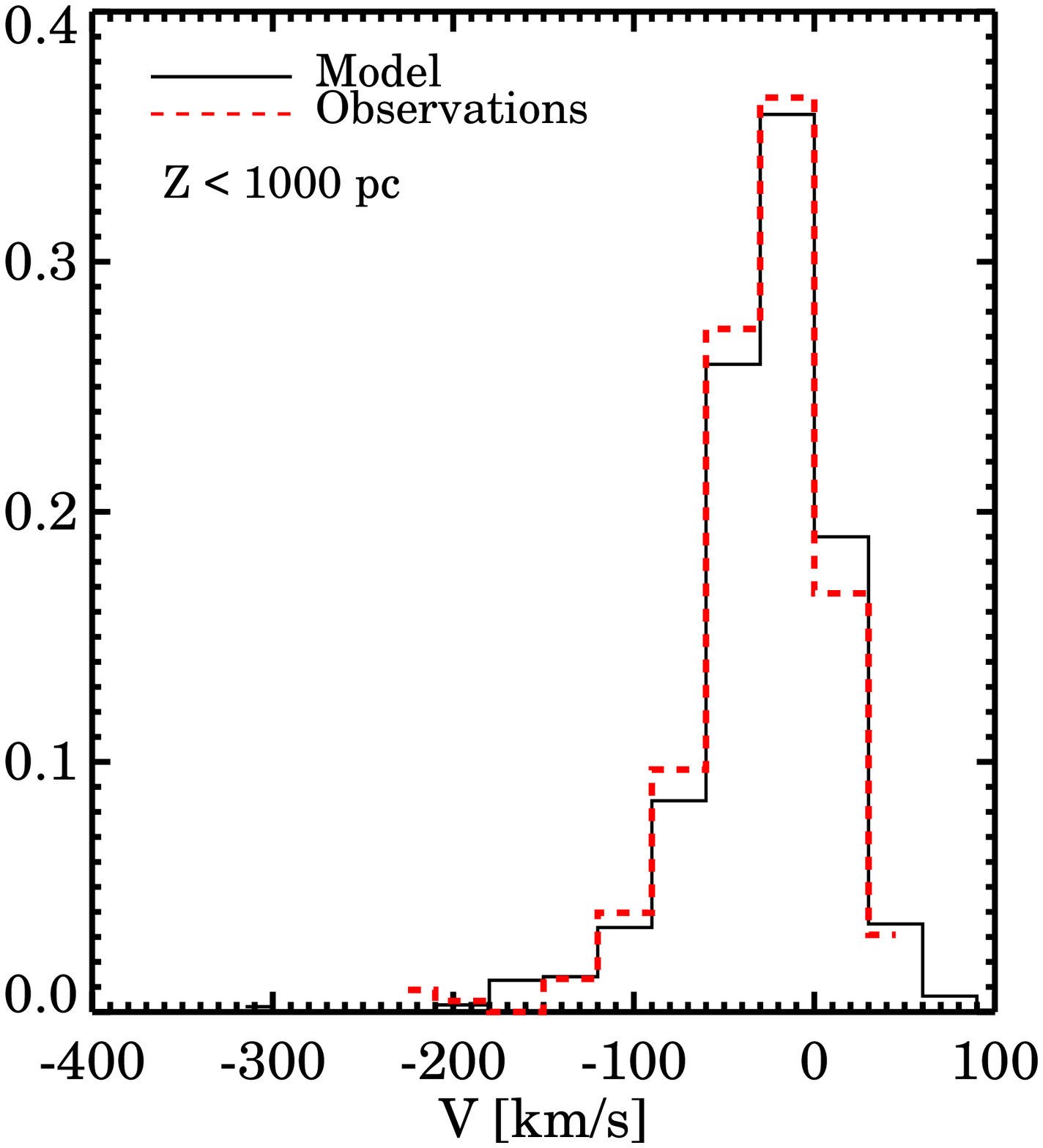}   & \includegraphics[width=5.6cm,height=5.6cm]{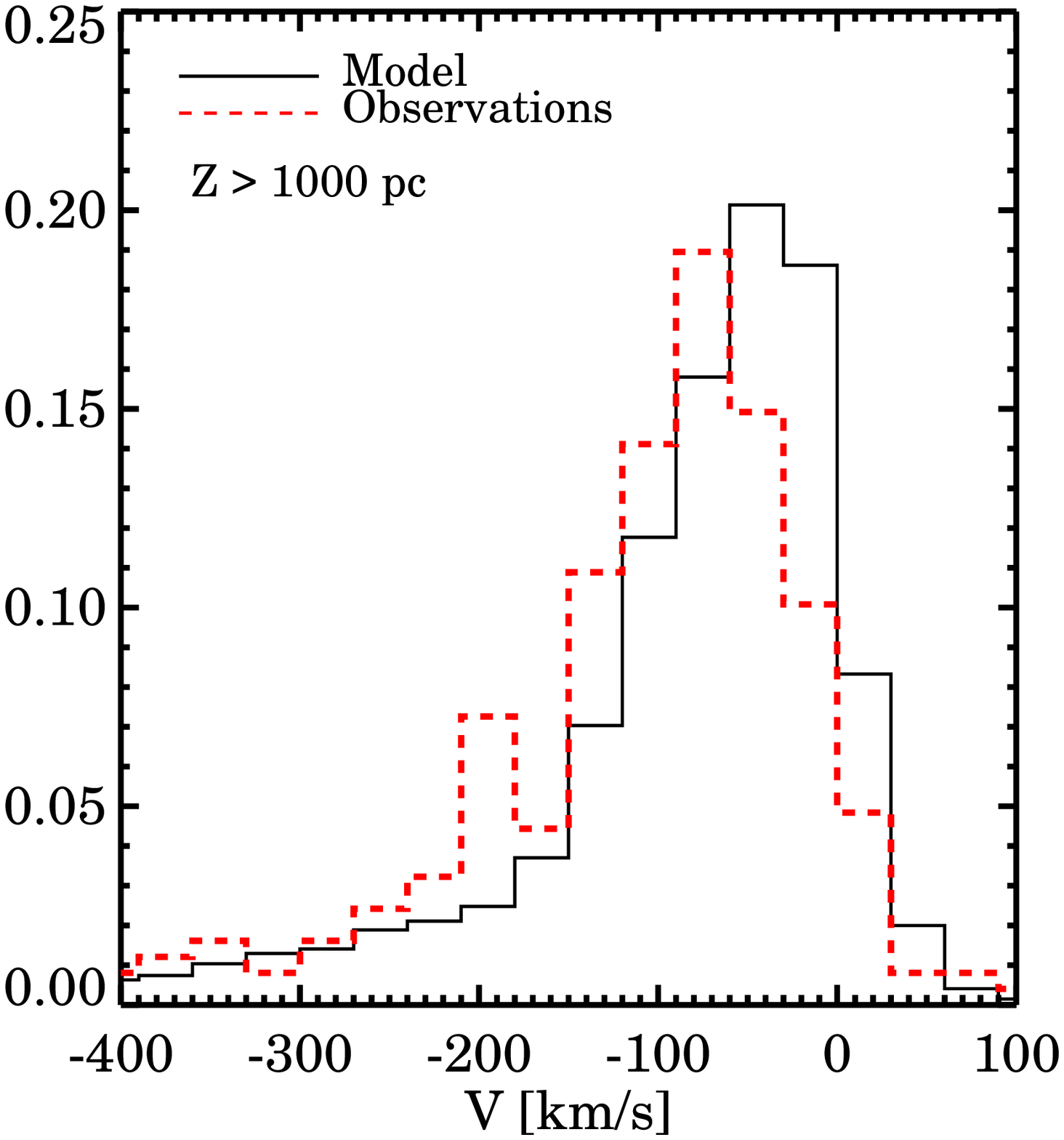} \\
      \includegraphics[width=5.6cm,height=5.6cm]{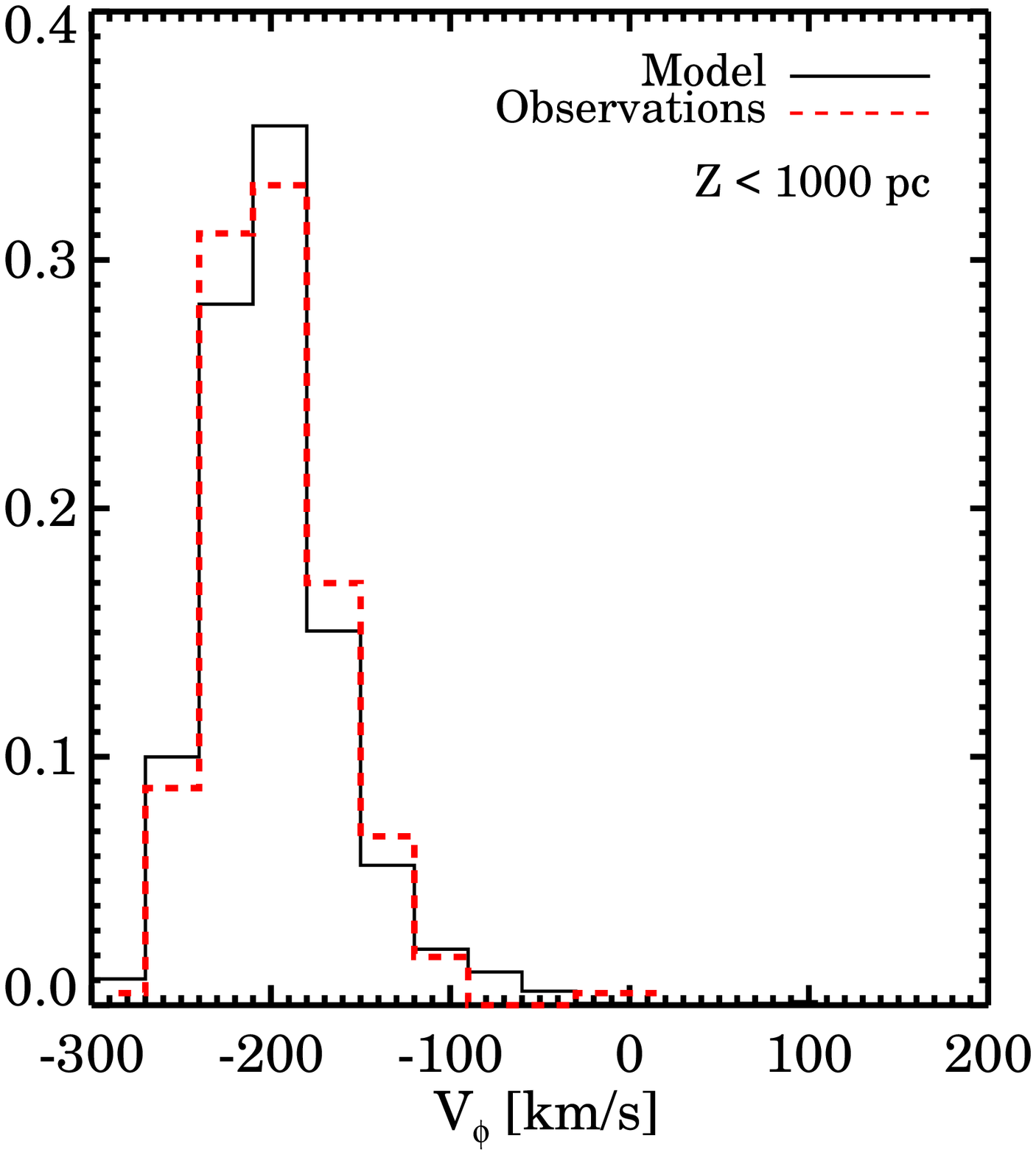}   & \includegraphics[width=5.6cm,height=5.6cm]{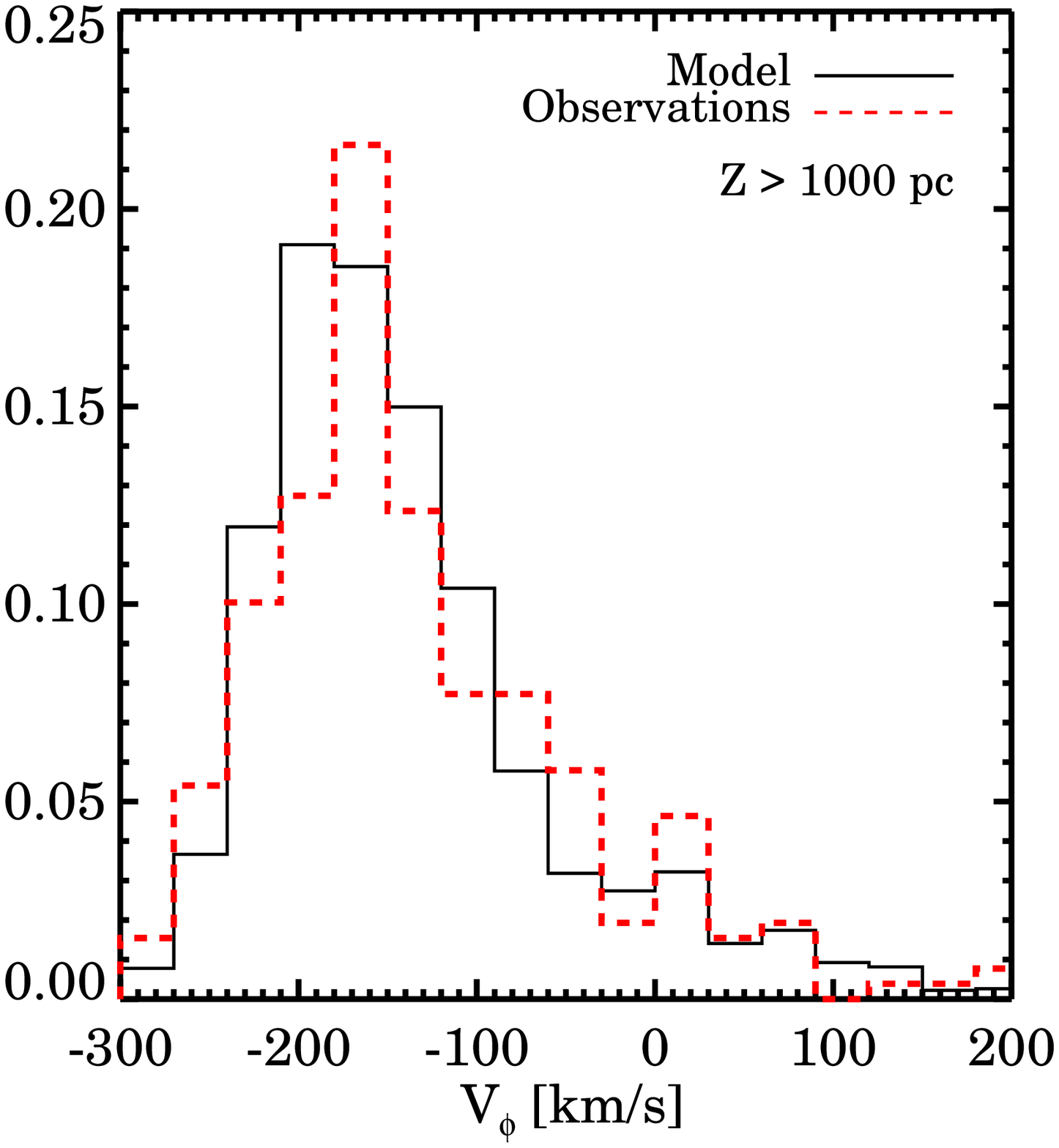} \\
      \includegraphics[width=5.6cm,height=5.6cm]{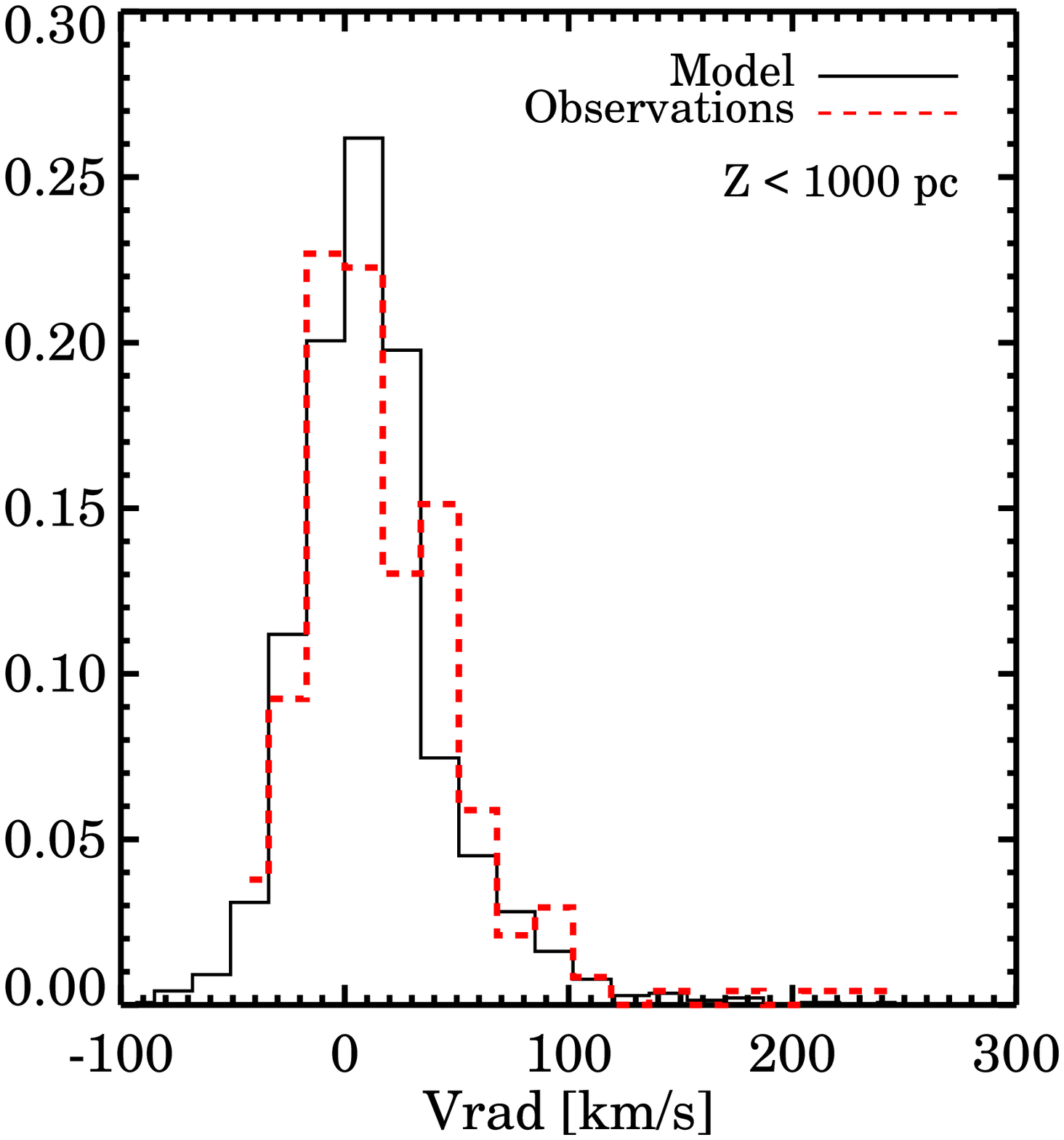}  &\includegraphics[width=5.6cm,height=5.6cm]{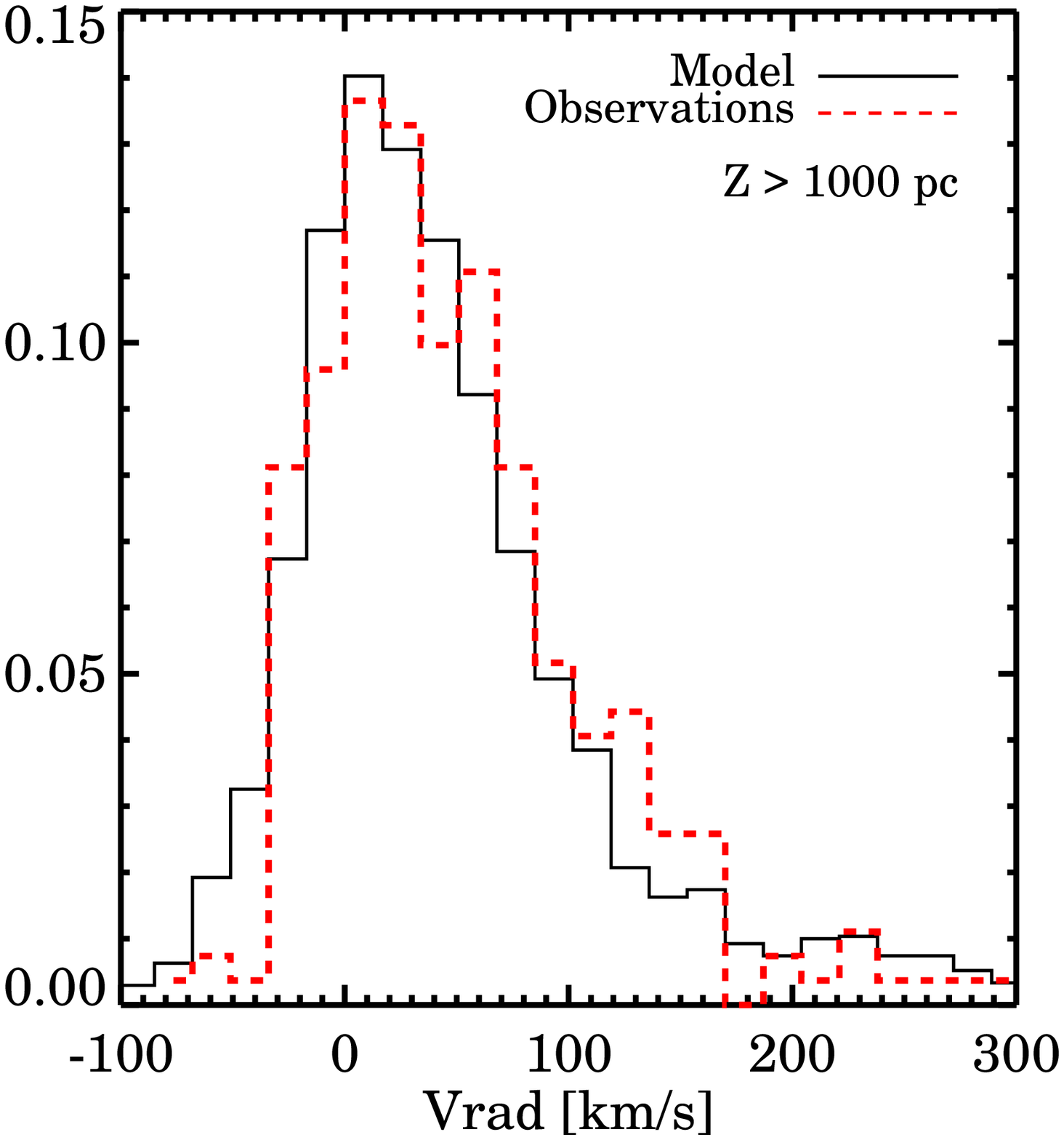}
    \end{tabular}
    \caption{Comparisons between our observations (red dashed histograms) and the Besan\c{c}on model predictions (black solid histograms) for the metallicities, rotational ($V$ in cartesian and $V_\phi$ in cylindrical coordinates system) and radial velocities. The left-side plots include the stars that are lower than 1~kpc, whereas the right-side plots include the stars farther than 1~kpc from the plane. Dotted black histogram corresponds to the raw Besan\c{c}on model (as downloaded from the web, but biased to match our magnitude distribution), whereas the  black continuous line corresponds to a model with a richer by 0.3 and 0.2~dex thick disc and halo, respectively.}
\label{fig:histograms_FLAMES}
  \end{center}
\end{figure*}

First, we divided the observed sample into stars lying closer and farther than 1~kpc from the galactic plane, with  201 and  251 targets, respectively.
 We recall that this cut at $Z=1$~kpc roughly corresponds to the scale height of the thick disc \citep{Juric_2008} and to the threshold height above which \citet{Gilmore_2002} identified a lagging population towards lower galactic latitudes ($b\sim 33^{\circ}$).

The left-side  histograms of Fig.~\ref{fig:histograms_FLAMES} represent a comparison between the model and the observations  for  metallicity, 
rotational velocity and  $V_\mathrm{rad}$ of the closest stars. They show that the model reproduces the observations close to the plane fairly well, except for the position of the metallicity peak, for which we find a more metal-poor distribution, by 0.15~dex.

On the other hand, for the more distant set (right-side histograms), several adjustments are needed to bring the metallicities and the velocities of the model to a  better agreement with the observations.
 Indeed,  the model considers a thick disc and a halo following Gaussian metallicity distributions, with a mean $\overline{[M/H]}_{TD}= -$0.78~dex, and 
$\overline{[M/H]}_{H}= -$1.78~dex, respectively. These values are too metal-poor compared to the mean thick disc and mean halo metallicities found in 
the literature \citep[see][]{Soubiran_2003, Fuhrmann_2008, Carollo_halo}. We find that a thick disc and a halo more metal-rich than the model default 
values by 0.3~dex and 0.2~dex respectively, lead to a much better agreement of the metallicity distribution, as can be seen from the 
comparison between the red dashed histogram (the observations) and the black one (the modified model) in the first row plots of 
Fig.~\ref{fig:histograms_FLAMES}. We therefore propose    
$\overline{[M/H]}_{TD}= -$0.48~dex, and   $\overline{[M/H]}_{H}= -$1.58~dex. We stress that the proposed value for the halo 
can suffer from the removal of some of the metal-poor stars, as described in Sect.~\ref{subsec:star_selection}.

As far as the velocity distributions are concerned, the model still nicely represents the U and W distributions far from the plane, but 
not the velocity in the direction of galactic rotation (V). Indeed, we find a shifted distribution with a peak around $-70$~\mbox{km s}$^{-1}$, 
lower by 20~\mbox{km s}$^{-1}$ compared to the predicted one.   This $\sim20$~\mbox{km s}$^{-1}$ difference is also seen for $V_\phi$, suggesting that its origin is not a spurious effect caused by the local, cartesian reference frame. 
Nevertheless, this effect is hardly visible from the radial velocity distribution, and the lag is much less clear compared to that visible 
in Fig.~2 of \citet{Gilmore_2002}. This might be partly because at higher latitudes the contribution of V to the $V_\mathrm{rad}$ 
is less important.

\begin{figure}
  \begin{center}
    \begin{tabular}{c}
      \includegraphics[width=7.5cm,height=7.2cm]{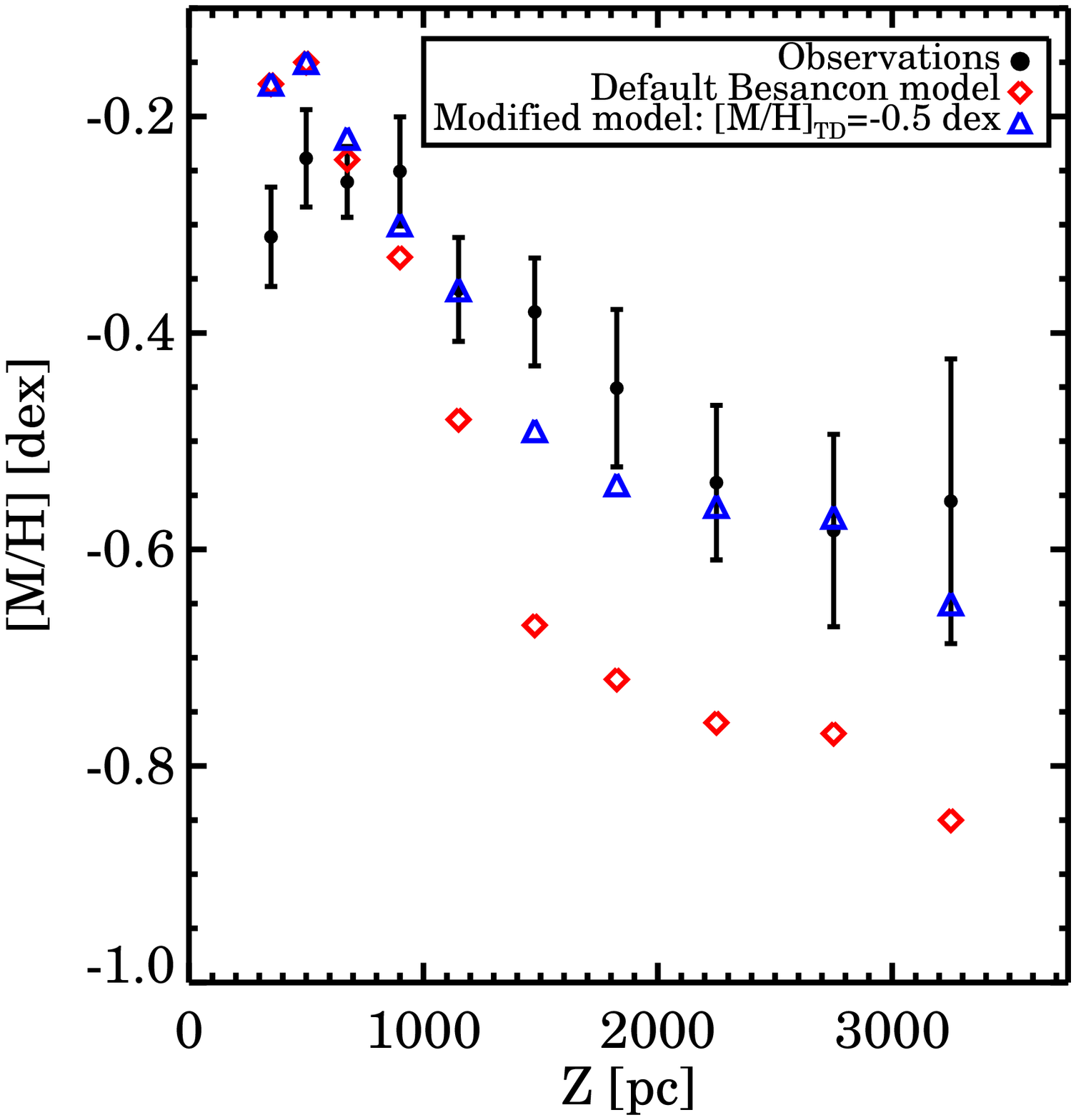} \\
      \includegraphics[width=7.5cm,height=7.2cm]{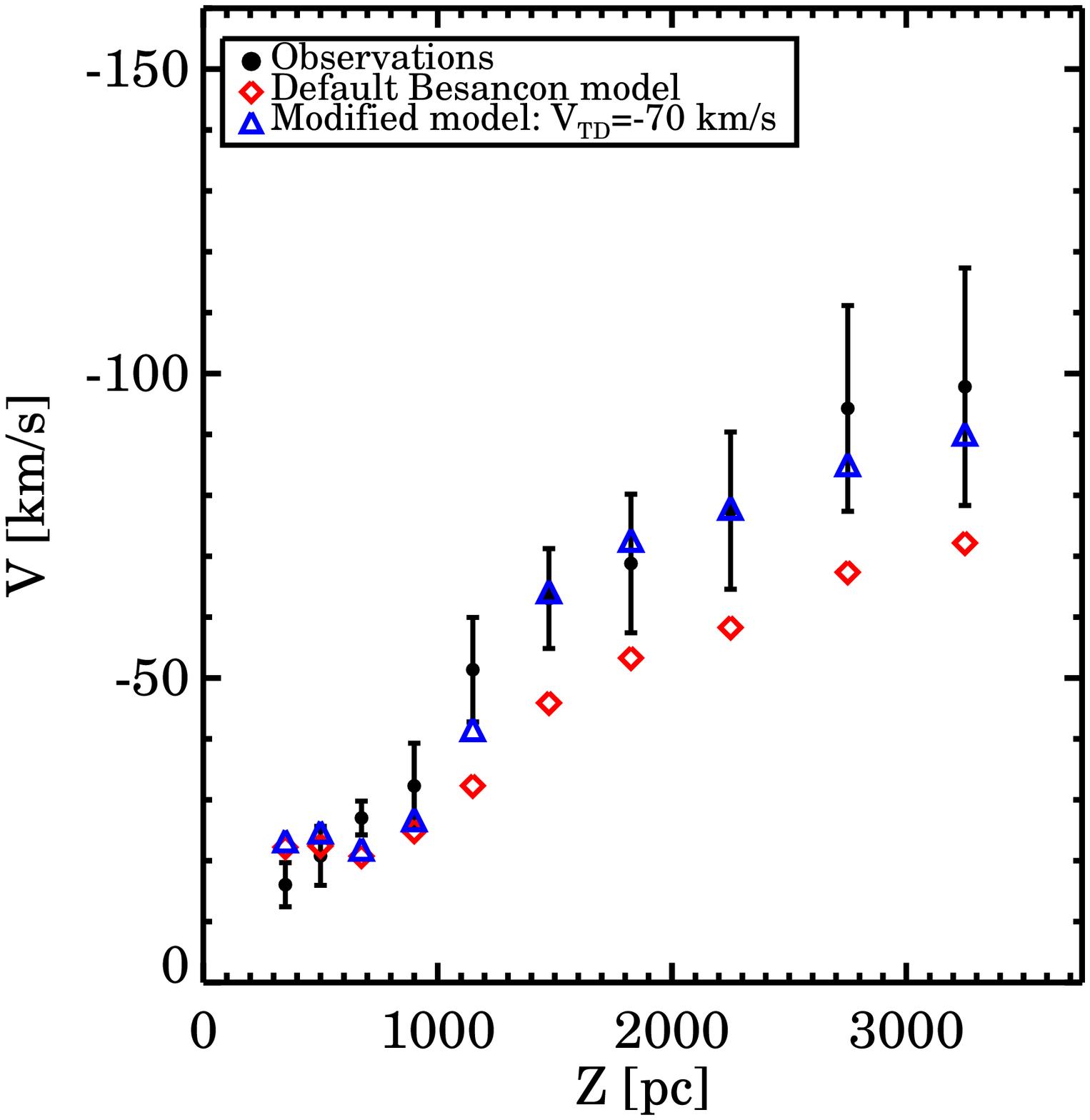} \\
      \includegraphics[width=7.5cm,height=7.2cm]{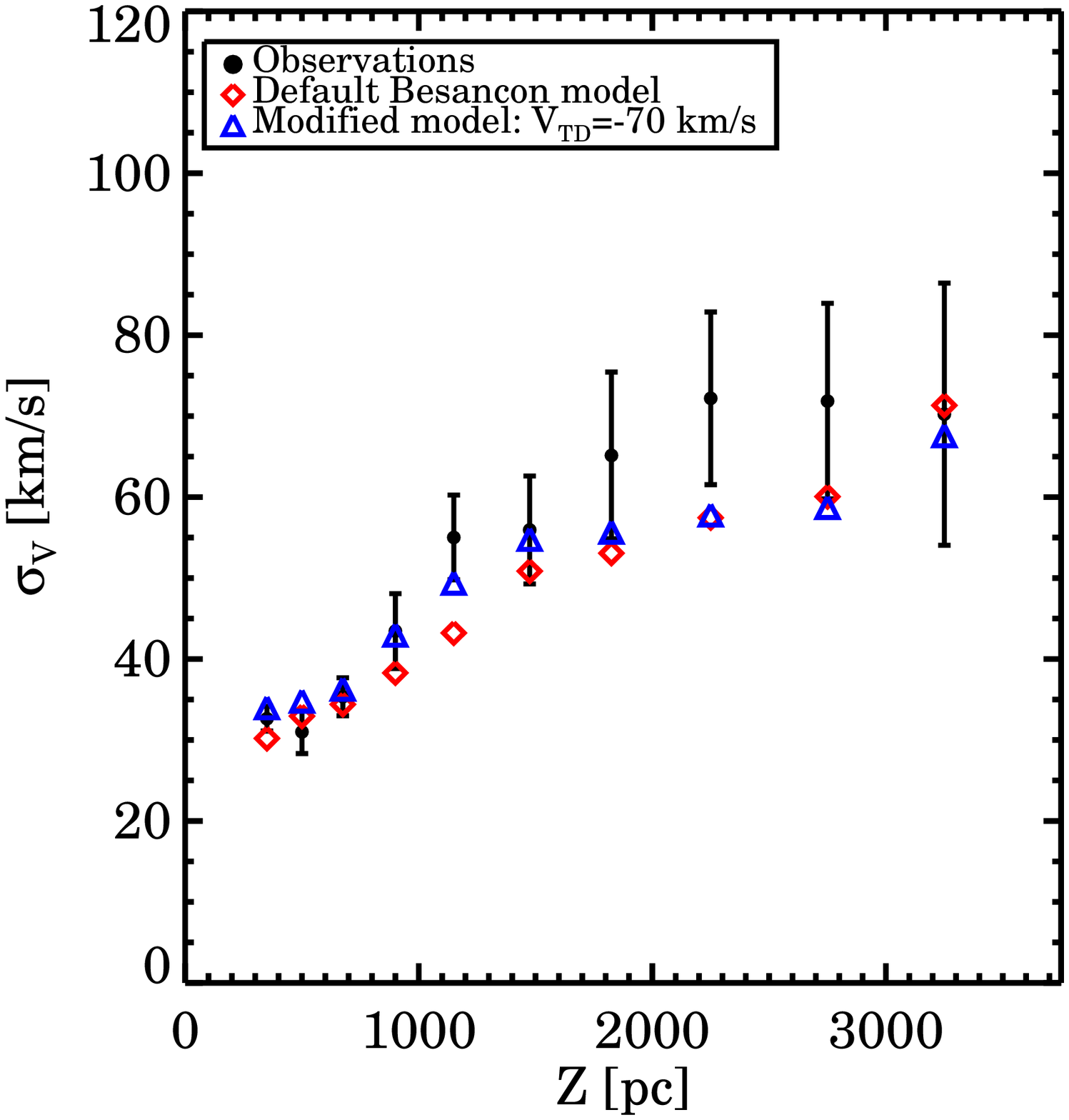} 
    \end{tabular}
\caption{Median metallicities and V-velocities at different height bins above the galactic plane. Each point has 1$\sigma$ uncertainties, obtained from $5\cdot10^3$ Monte-Carlo realisations on the position of the stars (and hence their velocities) and their metallicities. Red diamonds represent the predictions of the raw Besan\c{c}on model, whereas the blue triangles represent a modified model with metal-richer thick disc and halo, and a thick disc that lags behind the LSR by $\overline{V}=-70$~\mbox{km s}$^{-1}$.}
\label{fig:V_gradient}
  \end{center}
\end{figure}

To better characterise the observed sample, we studied the metallicity and the velocity distributions for narrower height bins  
up to 4~kpc from the plane. The size of the bins was chosen to include at least 20 stars.
To fully take into account the uncertainties in the computed positions and metallicities of the observed stars, the plots of 
Fig.~\ref{fig:V_gradient} were obtained
from $5\cdot10^3$ Monte-Carlo realisations on both parameters ({\it i.e.} $D$ and [M/H]). For each realisation we computed the new velocities  
and measured the median metallicity and the median V of the stars inside each bin. The mean value and the 1$\sigma$ uncertainties rising 
from the Monte-Carlo realisations were plotted. Finally, $\sigma_v$ was obtained by computing the mean robust 
deviation of V  inside each bin. 

A clear change of regime is observed in Fig.~\ref{fig:V_gradient} for [M/H], V and $\sigma_V$ around 1~kpc, {\it i.e.} where the 
transition between the thin and the thick disc is expected. 
Closer than 1~kpc the metallicity trend is flat,  whereas for the more distant stars a gradient 
\MetaZ\ is measured. This gradient, agrees (within the errors) 
with that found by \citet{Katz2011} and \citet{Ruchti11}. Yet this gradient might be under-estimated because as noticed 
previously, our sample might be lacking low-metallicity stars at long distances. 
Velocity gradients of \VphiZ\ and \SigmaVphiZ\  
%($\partial V_\phi/ \partial Z=18 \pm 7$~\mbox{km s}$^{-1}$~\mbox{kpc}$^{-1}$ , 
%$\partial \sigma_{V_\phi}/ \partial Z=16 \pm 6$~\mbox{km s}$^{-1}$\mbox{kpc}$^{-1}$) 
are observed for the stars between 1 and 4 kpc,  which agrees relatively well with the results found by \citet{Casetti11} and \citet{Girard_2006}. 

To better understand the origin of these trends, we used once more the Besan\c{c}on model and compared it to our observations. 
As suspected from the study above and  below 1~kpc, the raw Besan\c{c}on model does not mimic our data correctly (Fig.~\ref{fig:V_gradient}, 
red diamonds). 
On the other hand, a more metal-rich thick disc and halo, and a thick disc that lags behind the LSR by $\overline{V}=-70~$\mbox{km s}$^{-1}$, as suggested previously, leads to a much better agreement between the observations and the predictions (blue triangles). 
In particular, the measured trends seem to be explained in our mock model as a smooth transition between the different galactic populations.

Nevertheless, even with these adjustments the mock model does not represent all results we obtained. In particular, bins around $Z=1$~kpc still clearly disagree, corresponding to the height where the thick disc is expected to become the dominant population. 
In addition, the plateau at the high-metallicity regime is not well modelled either, suggesting a local density of the thick disc higher than 
the one assumed for the model,  correlated perhaps with a different scale height. Indeed, the model essentially predicts the number of thick
disc stars seen above one scale height, and thus the adopted thick disc scale height and $Z=0$ normalisation are degenerate.

Moreover, a clear correlation between $V_\phi$ and [M/H] is found for the stars between 0.8 and 2~kpc (\VphiMetaCorrel), in agreement with \citet{Spagna_2010} and \citet{Lee11}), though in disagreement with the SDSS view of \citet{Ivezic_2008}, based on photometric metallicities. We recall that according to the radial migration scenarios \citep{Roskar_2008, Radial_mixing_2009}, no or only a very small correlation is expected in the transition region, because the older stars that compose the thick disc have been radially well mixed.

%
%________________________________________________________________________
\subsection{Characterisation of the galactic components}
\label{subsec:components}
We now aim to select the thin disc, thick disc and halo members  to characterise these three galactic components.
We recall that there is no obvious predetermined way to define a sample of a purely single galactic component. 
Any attempt will produce samples contaminated by the other populations.

We used the kinematic approach of  \citet{Soubiran_Girard_2005, Bensby_2006,Ruchti_2010} 
to select the stars belonging to each galactic component (hereafter also called probabilistic approach). 
We recall that the advantage of this method is that no scale height is assumed for any population. 
The inconvenience is that it assumes a velocity ellipsoid, forcing in that way to find 
member candidates within the given dispersions (introducing for example  biases in the eccentricity distributions because
 the cold stars like thin disc's ones will have low eccentricities and very hot stars vey high eccentricities). 
This may lead to systematic misclassifications where the velocity distributions
overlap, or if the assumed distributions are not the expected ones (\textit{i.e.} non Gaussian or with significantly 
different means and/or dispersions).  

In practice, for each set of U, V and  W a membership probability is computed according to the following equation:
\begin{equation}
P=\frac{1}{(2\pi)^{3/2}  \sigma_U  \sigma_V  \sigma_W}\cdot \mathrm{exp} \left( -\frac{U^2}{2\sigma_U^2} 
- \frac{(V-V_\mathrm{lag})^2}{2\sigma_V^2} - \frac{W^2}{2\sigma_W^2} \right) .
\end{equation}
The adopted values of $\sigma_U$, $\sigma_V$, $\sigma_W$ and $V_\mathrm{lag}$ are the ones represented in Table~\ref{tab:char_Gal_components}. 
They are taken from \citet{Bensby_2006}, except for the $V_\mathrm{lag}$ of the thick disc, in which case we used the value suggested in Sect.~\ref{subsec:comparaison_besancon} of $-70$~\mbox{km s}$^{-1}$. 
The probability of belonging to one of the components has to be significantly higher than the probability of belonging to the others, to assign a target to it. 
In addition, one has to take into account the uncertainties on the measured velocities, to avoid massive misclassifications. 
In practice, instead of assigning a component based on the final values of U, V and W (presented in the online Table~\ref{online2}), 
we performed it for each of the $5\cdot10^3$ Monte-Carlo realisations used to compute these values (see Sect.~\ref{subsec:kinematics}). 
The adopted threshold in probability ratio above which a realisation is assigned to a component was fixed to four (we checked 
that our results are not affected by the adopted threshold, see below). The final assignment was then made by selecting the component 
which occurred in most realisations.

%In order to take into account the measurement errors, we assigned a membership for each of the $5 \cdot 10^3$ Monte-Carlo realisations
% of $D$, $\mu_l$, $\mu_b$ and $V_\mathrm{rad}$, done to determine the U, V, W velocities of each star (see Sect.~\ref{subsec:kinematics}). 
%The final assignment was then determined by selecting the component which occurred in majority.

\addtocounter {table} {1}
\begin{table}
\centering
\caption{Adopted values for the characterisation of the galactic components }
\begin{tabular}{lcccc}
\hline
\hline
Galactic     & $\sigma_U$ &  $\sigma_V$ &  $\sigma_W$ &  $V_\mathrm{lag}$ \\
component    & (\mbox{km s}$^{-1}$) &  (\mbox{km s}$^{-1}$) &  (\mbox{km s}$^{-1}$) & (\mbox{km s}$^{-1}$)  \\\hline
Thin disc    & 35  & 20  & 16   & -15 \\
Thick disc   & 67  & 38  & 35   & -70 \\ 
Halo         & 160 & 90  & 90   & -220 \\ \hline
\end{tabular}
\label{tab:char_Gal_components}
\end{table}

\begin{table}
\centering
\caption{Galactic population identification obtained from synthetic spectra of the Besan\c{c}on sample simulated with a
signal-to-noise ratio of approximately 20~pixel$^{-1}$}
\begin{tabular}{cc|cc|cc}
\hline
\hline
 \multicolumn{2}{c|}{thin disc}     & \multicolumn{2}{c|}{thick disc} & \multicolumn{2}{c}{halo}        \\ \hline
  (1)    & (2)     & (1)   & (2)     & (1)   & (2)  \\ \hline
 84\% D  &  77\% D   & 25\% D  &  15\% D   & 1\%  D  & 0\% D     \\ 
 16\% TD &  23\% TD  & 73\% TD &  78\% TD  & 56\% TD & 34\% TD     \\ 
 0\%  H  &  0\%  H   & 2\%  H  &  7\%  H   & 43\% H  & 66\% H     \\ \hline

\end{tabular}
\label{tab:identification}
\tablefoot{(1)~Selection according to velocities of the stars (probabilistic approach). (2)~Selection according to $Z$-distance. D, TD and H, correspond to thin disc, thick disc and halo stars, respectively, as identified by the Besan\c{c}on model.}
\end{table}

We tested this approach with the parameters obtained from the synthetic spectra of the Besan\c{c}on sample, presented in Sect.~\ref{subsec:distances}. The adopted  velocity ellipsoids are the same as in Table~\ref{tab:char_Gal_components}, but in that particular case, we chose a $V_\mathrm{lag}$ compatible with the model, of $-46$~\mbox{km s}$^{-1}$. 
Results for spectra with S/N$\sim$20$~\rm{pixel}^{-1}$ are shown in Table~\ref{tab:identification}, where we checked the true membership of each assigned star (labelled (1)). This approach identifies the thin disc targets quite well (first column), but 25\% of the thick disc candidates are actually thin disc members (third column). Furthermore, the majority of the identified  halo stars are, in fact,  thick disc members. 

The assignment to a component for most of the stars is independent of the adopted probability threshold 
(in our case: four). 
Indeed, for the majority of the stars more than half of the Monte-Carlo realisations result in the same assignment. Increasing the probability 
threshold will only increase the realisations for which the assignment will not be obtained. Hence, the component which will have occurred in 
the majority of realisations will remain unchanged, keeping that way the same membership assignment.
Finally, we stress that the results obtained when replacing the velocity ellipsoids of \citet{Bensby_2006} with those 
of \citet{Soubiran_2003} or \citet{Carollo_halo} remained identical. Indeed, the  differences in the velocity ellipsoids or the 
rotational lag are not different enough (few \mbox{km s}$^{-1}$) to introduce changes in the final candidate selection.

To obtain a  sample of thick disc and halo stars as pure as possible, we found that a candidate selection based on the distance 
above the galactic plane was 
preferred \citep[like in][for example]{Eccentricities_SDSS}. In this case, one has to adopt some values for the scale heights of each population, 
and a normalisation factor to estimate the pollution of the other components. Scale heights are a matter of debate, 
especially for the thick disc and the halo, but it is generally admitted that stars lying farther than $\sim$1-2 kpc and closer 
than $\sim$4-5 kpc from the galactic plane are thick disc dominated \citep{Siegel_2002, Juric_2008,DeJong_2010}.  Results obtained for the 
same Besan\c{c}on sample with this distance selection are also shown in Table~\ref{tab:identification} under the label (2). In that case, 
only 15\% of the thick disc stars are misclassified as thin disc members. In addition, the ratio of recovered halo targets has increased 
from 43\% to 66\% compared to the probabilistic approach.

Keeping these results in mind, we decided below, to investigate the results obtained by both methods ($Z$ selection and probabilistic approach) for our observed data.
 We performed Gaussian fits of the distributions of U, V, W and [M/H] for each  
 galactic component, and we  discuss the mean values and dispersions for each method. 
Results and their 1$\sigma$ uncertainties are shown in Table~\ref{tab:Galactic_components_means}.

The probabilistic approach assigned  154 stars to the thin disc, 193 stars to the thick disc and 105 stars to the halo. 
On the other hand, in the  $Z$ selection the  163 stars lying closer than $Z=800$~pc were assigned to the thin disc, the 187 stars lying between $1\leq Z \leq 4$~kpc to the thick disc, and the  45 stars above 5~kpc from the plane to the halo.

%
%__________________________________________________________________
\subsubsection{The thin disc}

The values found with the kinematic approach shown in Table~\ref{tab:Galactic_components_means} are consistent with the properties of the old thin disc \citep[][and references therein]{Vallenari2006, Soubiran_2003}. The mean eccentricity for the thin disc stars is  0.14, with a dispersion of 0.06 (Fig.~\ref{fig:eccentricities}), though  
we remind the reader that these values are expected to be greater in reality. Indeed, the high eccentricity tail of the thin disc cannot be selected with the adopted procedure because of the biases introduced to the orbital parameters by the kinematic selection.

The most distant thin disc star is found at Z$\sim 1874 \pm 224$~pc. We found one thin disc candidate with 
[M/H]$\sim-1.5$~dex, at $Z\sim250$~pc.  This star has $V_\mathrm{rad}=1.3~\pm~4.5$~\mbox{km s}$^{-1}$, which is typical of the thin disc.  Nevertheless, we rather suspect this target to be thick 
disc member, whose kinematics occupy the wings of the thick disc distribution function in a region that overlaps the thin disc 
distribution function. This nicely illustrates the limitations of our adopted Gaussian model for the distribution functions. 
A measurement of its $\alpha$-elements abundances would increase the dimensionality available for population classification, and so might 
help to identify its membership with confidence. \\

A selection according to $Z$-distance results in a hotter velocity ellipsoid and a lower mean metallicity compared to the kinematic selection. 
The higher lag and the lower metallicity found with this method is probably caused by the pollution from the thick disc. 
This is also suggested from the eccentricity distribution of Fig.~\ref{fig:eccentricities_Zselection}, where one can 
notice that the thin disc has an anomalously high number of stars with $\epsilon\gtrsim0.2$.
Indeed, we recall that in that case, stars up to $Z=800$~pc are considered to be thin disc members. 
This model is of course dynamically unphysical, and is adopted here merely as a limiting and convenient case 
to illustrate different population classification outcomes.
The contamination from the thick disc will thus depend mainly on the local density of the latter, whose values are found to vary in the literature from 2\%  up to 12\% \citep[see][for a review of the normalisation factors]{Arnadottir_2009}.

Considering a thin disc with $\overline{V_\phi}=-211$~\mbox{km s}$^{-1}$ (as found with the kinematic approach) and a 
thick disc with $\overline{V_\phi}=-166$~\mbox{km s}$^{-1}$, the contamination caused by the latter should be $\sim$19\% to recover the lag of $\sim-15$~\mbox{km s}$^{-1}$ that we measure.
Roughly the same result (18\%) is obtained when looking for the amount of thick disc contaminators which is needed to pass from $\overline{[M/H]}=-0.22$~dex (kinematic approach) to $-0.27$~dex ($Z$ selection) with a canonical thick disc metallicity of $-0.5$~dex. 

Because the kinematics are reasonably well established in the solar neighbourhood, we conclude for the thin disc that the probabilistic approach 
should return more robust results than the $Z$-distance selection.

%
%__________________________________________________________________
\subsubsection{The thick disc}
\label{subsec:thick_disc}

The mean $V$-velocity ($\overline{V}=-63\pm2$~\mbox{km s}$^{-1}$, $\overline{V_\phi}=-166\pm2$~\mbox{km s}$^{-1}$, see  Tables~\ref{tab:Galactic_components_means} and \ref{tab:Galactic_components_means02}) found with the kinematic approach  
is slightly higher than the typical canonical thick disc value, which less than $-50$~\mbox{km s}$^{-1}$ \citep[see, for example,][]{Wilson_RAVE_eccentricities}. 
The mean eccentricity is 0.33, with a dispersion of 0.13 (see Fig.~\ref{fig:eccentricities}). The closest thick disc star is found at
 $Z=129\pm10$~pc, and the  most distant one at Z$\sim 5.33 \pm 1.17$~kpc. We find that the metallicity of the thick disc extends 
from $\sim-1.8\pm0.1$~dex up to super-solar values of $\sim+0.25\pm0.1$~dex, and that $\overline{[M/H]}=-0.41\pm0.02$~dex.  Let us note  again that
the mean metallicity might be overestimated owing to the contribution of the thin disc (misclassifications for the stars lying
in the  low velocity tails), and that the kinematic selection introduces biases in the eccentricity distribution.   

On the other hand, the $Z$ selection results in  $\overline{V_R}$, $\overline{V_\phi}$, $\overline{V_Z}$ velocity distributions which are compatible with those found probabilistically\footnote{We note though that we found non-zero values for $\overline{V_R}$ and $\overline{V_Z}$. We searched for possible reasons for these offsets without success. These mean values could be possibly caused by small zero-point errors on the proper motions.}.  
Nevertheless, a slightly lower metallicity ($\overline{[M/H]}=-0.45$~dex) is found, which, interestingly, 
agrees well with the results suggested in Sect.~\ref{subsec:comparaison_besancon}, where we compared our results 
with those of the Besan\c{c}on model.

\begin{figure}
  \resizebox{\hsize}{!}{\includegraphics{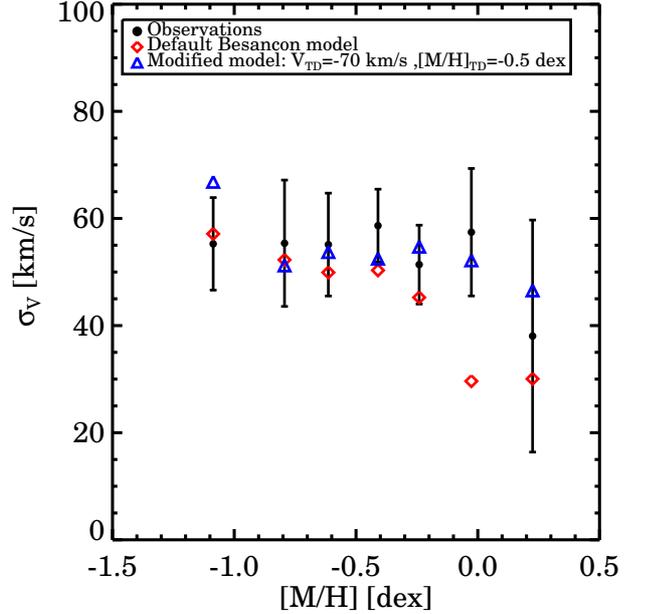}}
  \caption{V-velocity dispersion for different metallicity bins for the stars lying between $1<Z<4$~kpc above the plane. Red diamonds represent the raw Besan\c{c}on model, whereas the blue triangles represent the model with our preferred model, which has a mean metallicity $-0.5$~dex, and a mean galactic rotational velocity $\overline{V}=-70$~\mbox{km s}$^{-1}$ for the thick disc.}
  \label{fig:SigmaV_FeH_gradient}
\end{figure}

\begin{figure}
  \resizebox{\hsize}{!}{\includegraphics{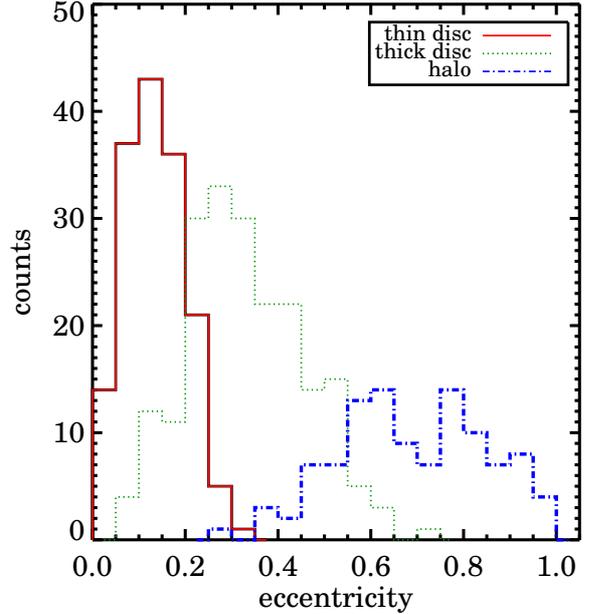}}
  \caption{Eccentricity distributions for the thin disc (in red), thick disc (in dotted green) and halo (in dotted-dashed blue). The candidates were selected 
  based on their kinematics, according to the  probabilistic approach of \citet{Ruchti_2010}. This selection therefore introduces biases in the eccentricity distributions because the tails of the velocity distributions cannot be selected.}
  \label{fig:eccentricities}
\end{figure}

The argument that the thick disc is a distinct population compared to the thin disc is even more accredited by the plot of $\sigma_V$ 
versus [M/H] (Fig.~\ref{fig:SigmaV_FeH_gradient}), obtained for the stars lying between 1 and 4~kpc. We see that for typical thick disc 
metallicities ($-1 <$[M/H]$<-0.2$~dex) the velocity dispersion is constant within the errors, around $\sigma_V \sim$55~\mbox{km s}$^{-1}$. 
At these heights and metallicities the pollution from the other components is expected to be very low (too high and too metal-poor for 
the thin disc, too low and too metal-rich for the halo). Hence, this plot seems to highlight the intrinsic velocity dispersion of the thick disc.   
But we cannot rule out the existence of an intrinsic vertical gradient in the metallicity and rotational velocity of the thick disc.
Higher resolution spectroscopy and/or higher number of targets like the forthcoming Gaia-ESO survey will help to answer this question, though. 
A cleaner thick disc sample might be obtained by separating the thin disc from the thick disc based on the [$\alpha$/Fe] ratio, and the 
halo from the thick disc by means of higher statistics. 

\begin{figure}
  \resizebox{\hsize}{!}{\includegraphics{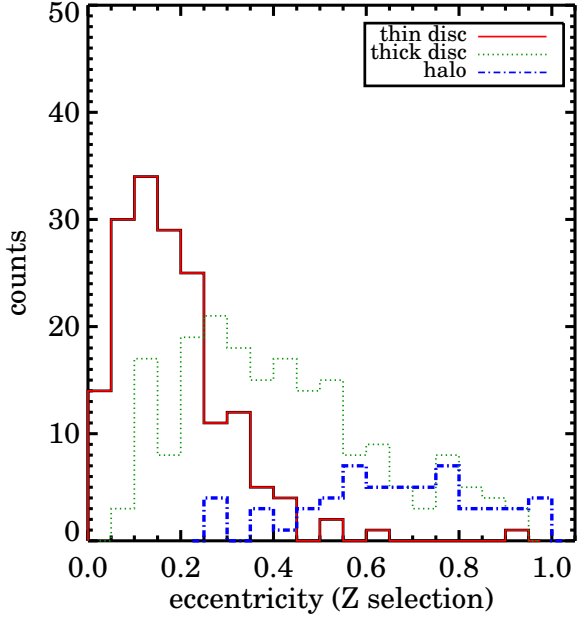}}
  \caption{Eccentricity distributions for the thin disc (in red), thick disc (in dotted green) and halo (in dotted-dashed blue). The candidates were selected 
  based on their distance above the galactic plane.}
  \label{fig:eccentricities_Zselection}
\end{figure}

Finally, the orbital eccentricities of the thick disc obtained with the two methods have to be discussed (see Fig.~\ref{fig:eccentricities} 
and Fig.~\ref{fig:eccentricities_Zselection}). 
Very many authors have recognised that   the shape of the eccentricity distribution can give valuable 
information about the formation mechanism of the thick disc and the accreted or not accreted origin of its stars \citep{Sales_2009, Di_Matteo_2011}.
Although for both selection methods the distribution tails can suffer from individual uncertainties or wrong population
assignments, the position of the peak is  reliable information according to these authors. 
The eccentricities in Fig.~\ref{fig:eccentricities} suffer from selection biases, as discussed in the previous sections, because 
the membership was decided on kinematic criteria. Hence, no conclusions should be taken based on these results.

On the other hand, Fig.~\ref{fig:eccentricities_Zselection} suggests a peak at rather low values, 
around $\epsilon\sim0.3$.  Hence, the pure accretion scenario, as in \citet{Galaxy_model_Abadi}, which supposes a 
peak around $\epsilon\sim0.5$, can be ruled out \citep[we note though that these results have been recently debated by ][]{Navarro_GrandBo}. 
Nevertheless, no separation can be made between the heating, migration 
or merger scenarios, which differ mainly in the detailed shape, especially in the breadth of the distribution. 

We note, however, that Fig.~\ref{fig:eccentricities_Zselection} suggests a broader distribution than has been found so far for
more local samples. This is very interesting because it might suggest that accretion has contributed slightly
more compared to local sample, and that the dominant mechanism for the formation of the thick disk
may well have varied with distance from the galactic center.

%_____________________________________________________________________
\subsubsection{The halo}
\label{subsec:halo_selection}
 
The addition of the individual errors on $V_\mathrm{rad}$, $\mu_l$, $\mu_b$ and $D$ leads to an unreliable selection of halo
stars based only on their kinematics. Indeed, we recall that our tests conducted on synthetic spectra of the  mock Besan\c{c}on catalogue
showed that a probabilistic assignment of halo membership inevitably generates a sample strongly polluted by the thick disc.   

To minimise this pollution, our tests suggest to perform a selection on $Z$ instead.
For the  45 stars being more distant from the plane than $Z=5$~kpc, the mean eccentricity is  0.69, with a dispersion of 0.16.
We find a mean metallicity of  $\overline{[M/H]}=-0.92\pm0.06$~dex, with $\sigma_{[M/H]}=0.76\pm0.06$~dex. 
These values are too metal-rich, 
compared to the commonly admitted value of $\sim-$1.6~dex for the inner halo \citep{Carollo_halo}. 
 In addition, we find mean $\overline{V_Z}$ velocities significantly different from zero, which is not expected. 
These odd values are likely caused by the large uncertainty expected 
for the most distant stars for both the distance and hence the velocities  to our star selection criterion, 
 which created a biased and unreliable halo sample (see Fig.\ref{fig:mv_vs_vrad}) and probably to an  underestimation of 
the errors for the proper motions for the most distant stars 
(we recall that we assumed a constant error of 2 mas/year).

This could also explain the fairly large velocity dispersions that we find compared to those expected for the halo. For example, using 
SEGUE spectra, \citet{Carollo_halo} found for the inner halo $(\sigma_{U}, \sigma_{V}, \sigma_W)=(150 \pm 2, 95 \pm 2, 85 \pm 1)$. 

We note, however,  that another possible solution to explain a part of this discrepancy can be found in \citet{Mizutani03}, who  suggested that 
the retrograde stars in \citet{Gilmore_2002} could be the debris of $\omega$~Cen.
 We searched among our retrograde stars for a possible identification of $\omega$~Cen candidates (e.g. in the metallicity space) without any success.
Therefore we favour small-number statistics to explain these large velocity dispersions, although we cannot exclude their presence.

\begin{table*}
\centering
\caption{Mean metallicity and kinematic values (not corrected for observational errors) of the different galactic components (see Sect. \ref{subsec:components}) in cylindrical coordinates.}
\label{tab:Galactic_components_means}

\begin{tabular}{lcccccc cccccc}
\hline
\hline
Galactic component  & N &$\overline{V_R}$ &  $\overline{V_\phi}$ &  $\overline{V_Z}$ &$\overline{[M/H]}$ & $\sigma_{V_R}$ & $\sigma_{V_\phi}$ & $\sigma_{V_Z}$ & $\sigma_\mathrm{[M/H]}$ \\
          & & (\mbox{km s}$^{-1}$) &  (\mbox{km s}$^{-1}$)  & (\mbox{km s}$^{-1}$) & (dex)& (\mbox{km s}$^{-1}$) &  (\mbox{km s}$^{-1}$)  & (\mbox{km s}$^{-1}$) & (dex) \\ \hline
Thin disc$_{kine}$  & 154 & $-8\pm2$   & $-211\pm1$   & $-5\pm1$   & $-0.22\pm0.02$ & $38\pm2$   & $26\pm1$   & $20\pm1$   & $0.28\pm0.02$ \\
Thick disc$_{kine}$ & 193 & $ 8\pm2$   & $-166\pm2$   & $-3\pm2$   & $-0.41\pm0.02$ & $56\pm4$   & $42\pm3$   & $55\pm2$   & $0.34\pm0.02$ \\
Inner halo$_{kine}$ & 105 & $38\pm13$  & $-60\pm10$   & $-23\pm10$ & $-0.70\pm0.03$ & $194\pm14$ &$121\pm13$  & $122\pm13$ & $0.59\pm0.05$ \\ \hline

Thin disc$_{Z}$  & 163 & $-1\pm1$  & $-204\pm1$   & $-2\pm1$   & $-0.27\pm0.02$ & $43\pm2$  & $33\pm1$   & $25\pm1$   & $0.31\pm0.02$ \\
Thick disc$_{Z}$ & 187 & $-2\pm3$  & $-167\pm3$   & $-12\pm2$  & $-0.45\pm0.02$ & $66\pm5$  & $57\pm4$   & $53\pm3$   & $0.36\pm0.02$ \\
Inner halo$_{Z}$ &  45 & $11\pm22$ & $-47\pm21$   & $-48\pm17$ & $-0.92\pm0.06$ & $215\pm31$&$166\pm28$  & $146\pm22$  & $0.76\pm0.06$ \\ \hline
\end{tabular}
\tablefoot{Thin disc$_{Z}$, thick disc$_{Z}$ and inner halo$_{Z}$ are defined as the stars lying below $Z< 800$~pc, between $1 < Z < 4$~kpc and 
above $Z>5$~kpc, respectively.  }
\end{table*}

\begin{table*}
\centering
\caption{Same as Table~\ref{tab:Galactic_components_means} but in cartesian coordinates.}
\label{tab:Galactic_components_means02}

\begin{tabular}{lcccccc cccccc}
\hline
\hline
Galactic component  &$\overline{U}$ &  $\overline{V}$ &  $\overline{W}$  & $\sigma_U$ & $\sigma_V$ & $\sigma_W$  \\
          &(\mbox{km s}$^{-1}$)  & (\mbox{km s}$^{-1}$) &  (\mbox{km s}$^{-1}$)  & (\mbox{km s}$^{-1}$) &(\mbox{km s}$^{-1}$) &  (\mbox{km s}$^{-1}$)   \\ \hline
Thin disc$_{kine}$  & $-18\pm2$   & $-14\pm1$   & $-5\pm1$   & $38\pm2$   & $25\pm2$   & $20\pm1$    \\
Thick disc$_{kine}$ & $-40\pm3$   & $-63\pm2$   & $-3\pm2$   & $58\pm4$   & $40\pm3$   & $55\pm2$   \\
Inner halo$_{kine}$ & $-65\pm13$  & $-175\pm9$  & $-23\pm10$ & $208\pm14$ &$97\pm12$   & $122\pm13$  \\ \hline

Thin disc$_{Z}$  & $-20\pm1$  & $-21\pm1$   & $-2\pm1$   & $43\pm2$  & $32\pm1$   & $25\pm1$    \\
Thick disc$_{Z}$ & $-34\pm4$  & $-69\pm3$   & $-14\pm2$  & $70\pm6$  & $55\pm4$   & $51\pm3$    \\
Inner halo$_{Z}$ & $-33\pm29$ & $-191\pm25$ & $-49\pm38$ & $223\pm39$&$142\pm28$  & $158\pm34$   \\ \hline
\end{tabular}
\tablefoot{Thin disc$_{Z}$, thick disc$_{Z}$ and inner halo$_{Z}$ are defined as in Table~\ref{tab:Galactic_components_means}. The mean velocities are given without taking into account the solar motions ($U_\odot, V_\odot, W_\odot$). }
\end{table*}

\begin{figure*}
  \begin{center}
    \begin{tabular}{ll}
      \includegraphics[width=6.cm,height=6.cm]{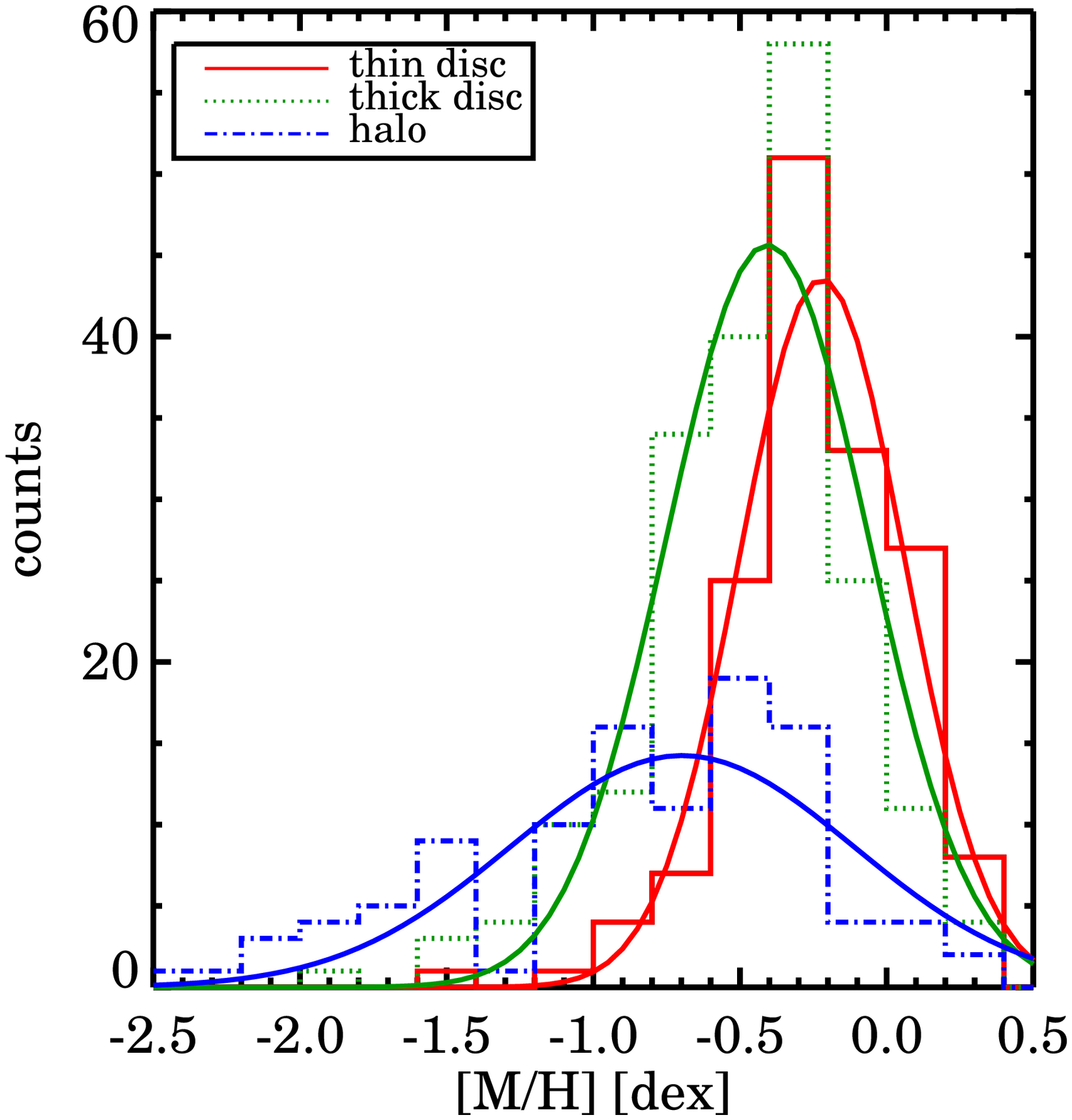} & \includegraphics[width=6cm,height=6cm]{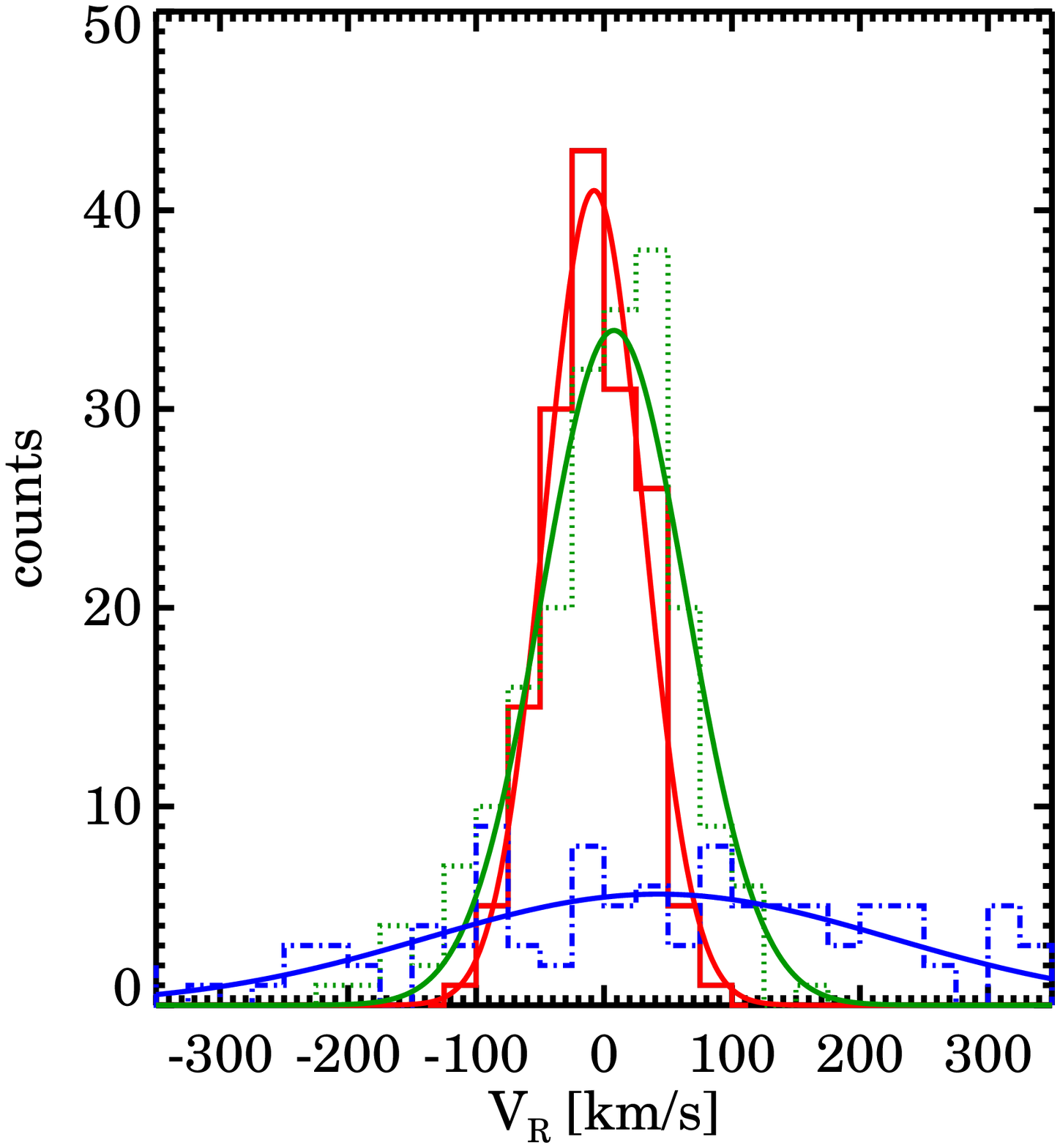} \\
      \includegraphics[width=6cm,height=6cm]{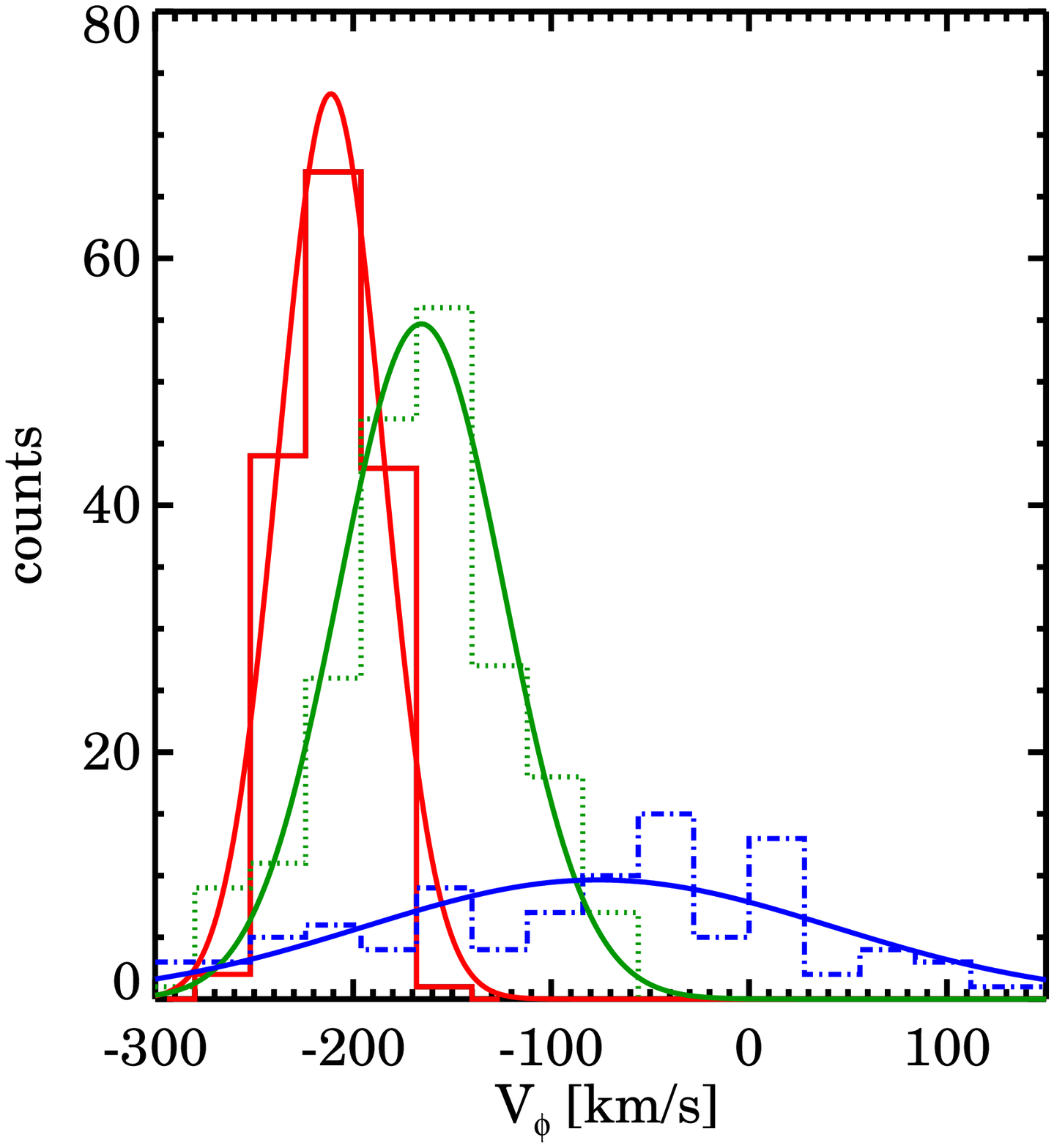} & \includegraphics[width=6cm,height=6cm]{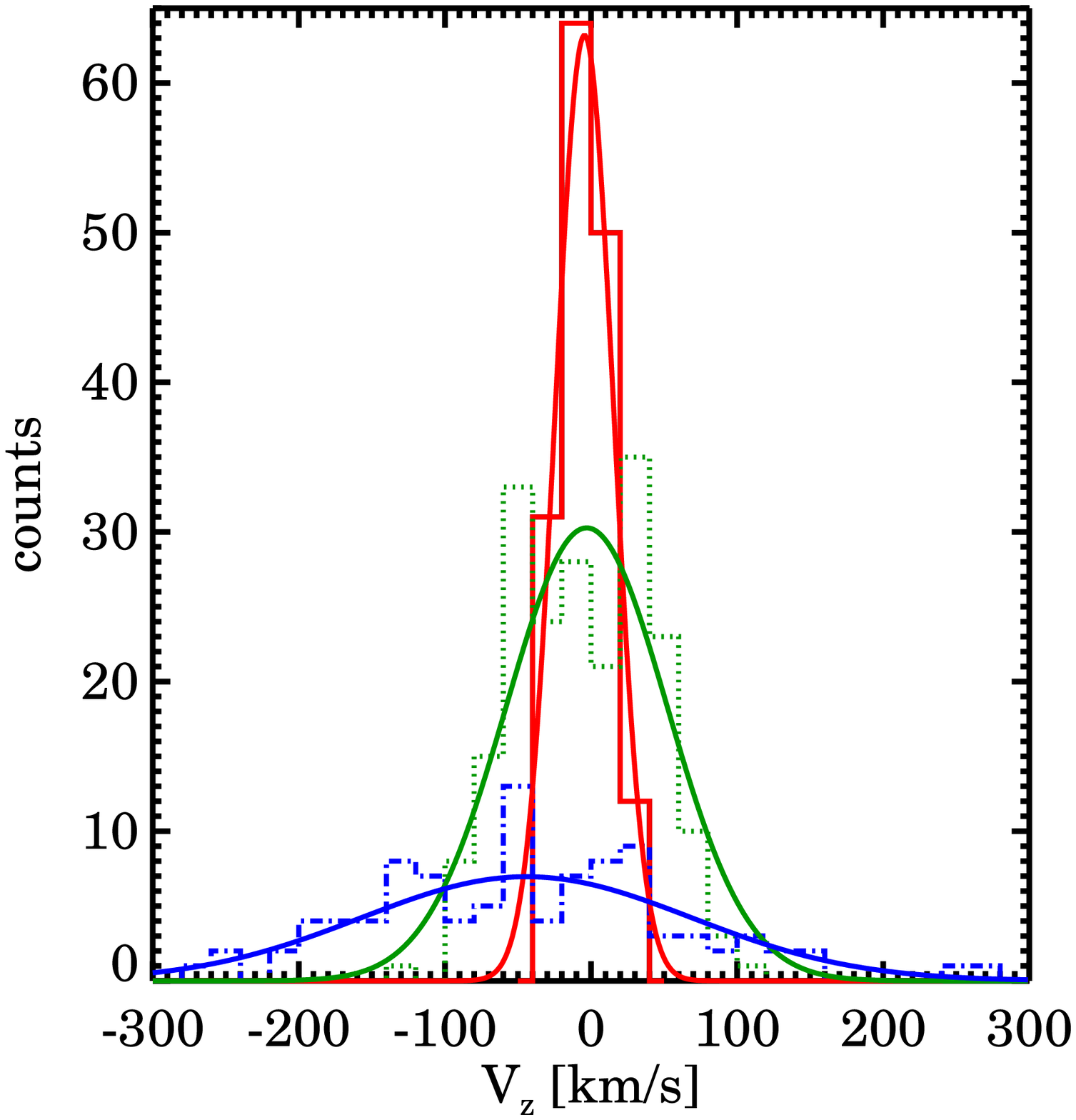} \\
    \end{tabular}
\caption{Metallicity and velocity distributions for the galactic components selected according to their position above the galactic plane. The Gaussian fits were obtained taking into account the errors on each parameter with $5 \cdot 10^3$ Monte-Carlo realisations.}
\label{fig:Gaussian_fits}
  \end{center}
\end{figure*}

%
%________________________________________________________________________________________________
\section{Derivation of the radial scale lengths and scale heights}
\label{sec:scale_lengths}
Supposing that the thick disc and the thin disc are in equilibrium, the velocity ellipsoids that were derived in the previous sections can be used with the Jeans equation to infer an estimate of their radial scale lengths ($h_R$) and  scale heights ($h_Z$). Below we considered as thin disc the stars below 800~pc from the galactic plane, and as thick disc the stars between 1 and 3~kpc to avoid a strong contamination from the other components. In addition, the dispersions of the velocity ellipsoids were corrected from the observational errors by taking them out quadratically, as in \cite{Jones88}. 
In cylindrical coordinates the radial and azimuthal components of the Jeans equation are 
\begin{eqnarray}
v_c^2-\overline{v_\phi}^2&=&\sigma^2_{V_R} \left( \frac{\sigma^2_{V_\phi}}{\sigma^2_{V_R}} -1 - \frac{\partial ln(\rho \sigma^2_{V_R})}{\partial ln R} - \frac{r}{\sigma^2_{V_R}} \frac{\partial \sigma^2_{V_{R,Z}}}{\partial Z} \right) 
\label{eqn:Jeans_general}\\
\rho K_Z &=& \frac{\partial \rho \sigma^2_{V_Z}}{\partial Z} + \frac{1}{R} \frac{\partial R \rho \sigma^2_{V_{R,Z}}}{\partial R} 
\label{eqn:vertical_jeans_general}
\end{eqnarray}
where $\rho$ is the density of the considered galactic component, $V_c=220$~\mbox{km s}$^{-1}$ is the circular velocity at the solar radius, $\overline{V_\phi}$ is the mean rotational velocity of the stars having the $\sigma_{V_R},\sigma_{V_\phi}, \sigma_{V_Z}$ velocity dispersions, $\sigma^2_{V_{R,Z}}=\overline{V_R V_Z}-\overline{V_R}~\overline{V_Z}$, and $K_Z$ is the vertical galactic acceleration.

\subsection{Radial scale lengths} 
We consider that $\rho(R) \propto \mathrm{exp}(-R/h_R)$, and that  $\sigma^2_{V_R}$ has the same radial dependence as $\rho$ \citep[as in][]{Carollo_halo}. Therefore, $\sigma^2_{V_R} \propto \mathrm{exp}(-R/h_R)$.
 In addition, one can assume that  $\partial \sigma^2_{V_{R,Z}}/\partial Z \approx 0$, which is true if the galactic potential is an infinite constant surface density sheet \citep{Gilmore89}. In this case, the axes of the velocity ellipsoid are aligned with the cylindrical coordinates, and  
Eq.~\ref{eqn:Jeans_general} becomes:
\begin{equation}
\frac{\sigma^2_{V_\phi}}{\sigma^2_{V_R}} -1 + \frac{2R}{h_R} - \frac{v_c^2-\overline{v_\phi}^2}{\sigma^2_{V_R}}=0 .
\label{eqn:jeans1}
\end{equation}
An alternative option is to consider that the galactic potential is  dominated by a centrally concentrated mass distribution and that the local velocity ellipsoid points towards the galactic centre \citep{Gilmore89,Siebert08}. In that case, the previous term becomes
\begin{equation}
\frac{r}{\sigma^2_{V_R}} \frac{\partial \sigma^2_{V_{R,Z}}}{\partial Z }\approx 1- \frac{\sigma^2_{V_Z}}{\sigma^2_{V_R}}
\label{eqn:approx_Gilmore}
\end{equation}
Equation \ref{eqn:Jeans_general} can then be rewritten as follows:
\begin{equation}
\frac{\sigma^2_{V_\phi}}{\sigma^2_{V_R}} -2 + \frac{2R}{h_R} - \frac{v_c^2-\overline{v_\phi}^2}{\sigma^2_{V_R}} + \frac{\sigma^2_{V_Z}}{\sigma^2_{V_R}}=0.
\label{eqn:jeans2}
\end{equation}

Each of the terms of Eq.~\ref{eqn:jeans1} or Eq.~\ref{eqn:jeans2} were measured in our data, leaving as the only free variable the radial scale length $h_R$ of the discs. 
With the values derived for the thick disc of ($\sigma_{V_R}; \sigma_{V_\phi}; \overline{V_\phi})=(66\pm 5; 57\pm 4; -167 \pm 3)$~\mbox{km s}$^{-1}$, we find 
$h_R=3.6 \pm 0.8$~kpc using Eq.~\ref{eqn:jeans1}, and  \HrValue \ using Eq.~\ref{eqn:jeans2}. These two values reasonably agree between them, and are found in the upper end of the values cited in the literature (ranging from 2.2~kpc \citep{Carollo_halo} up to 3.6~kpc \citep{Juric_2008}, or even 4.5~kpc in the case of \citet{Chiba_and_Beers}).  

Using Eq.~\ref{eqn:jeans2} and ($\sigma_{V_R}; \sigma_{V_\phi}; \overline{V_\phi})=(43\pm 2; 33\pm 1; -204 \pm 1)$~\mbox{km s}$^{-1}$, we find that the thin disc  has a similar radial extent  within our uncertainties as the thick disc, with $h_R=2.9 \pm 0.2$~kpc. 
A smaller thin disc has been suggested by other recent observations \citep[see][]{Juric_2008}, but once more, the value we derived is at the upper end of the previously reported values in the literature. 
Still, an extended thin disc like this is plausible because our data mainly probe the old thin disc, which is likely to be more extended than its younger counterpart.

\subsection{Scale heights}
We assume  that the last term of Eq.~\ref{eqn:vertical_jeans_general} is negligible, because we are far from the galactic centre, and that $\rho(Z) \propto \mathrm{exp}(-Z/h_Z)$. Equation~\ref{eqn:vertical_jeans_general}  hence becomes
\begin{equation}
\frac{\partial ln \sigma^2_{V_Z}}{\partial Z}- \frac{1}{h_Z}+\frac{K_Z}{\sigma^2_{V_Z}}=0.
\label{eqn:vertical_jeans}
\end{equation}
We used $K_Z=2\pi G \times 71~M_\odot pc^{-2}$ derived by \citet{Kuijken91} at $|Z|=1.1$~kpc, but we note  that this value of $K_Z$ might differ at the distances where our targets are observed.
We also used for the thick disc the value derived from our data of $\partial \sigma_{V_Z} / \partial Z = 15 \pm 7$~\mbox{km s}$^{-1}$~\mbox{kpc}$^{-1}$ and $\sigma_{V_Z}=53 \pm 3$~\mbox{km s}$^{-1}$. Hence, for the thick disc, we find \HzValue.

 We found for the thin disc that  $\partial \sigma_{V_Z} / \partial Z = 19 \pm 10$~\mbox{km s}$^{-1}$~\mbox{kpc}$^{-1}$ and $\sigma_{V_Z}=25 \pm 1$~\mbox{km s}$^{-1}$, resulting in $h_Z=216 \pm 13$~pc.

The derived values for both components agree well with, for example,  \citet{Juric_2008}, who suggested a thin disc with $h_Z=300$~pc, and a thick disc with $h_Z=900$~pc.

\subsection{Metallicity dependence of the scale lengths of the thick disc}
To investigate furthermore the metallicity gradients found for the rotational velocity of the thick disc, we computed the radial scale lengths and  scale heights for different metallicity bins using Eq.~\ref{eqn:jeans2} and Eq.~\ref{eqn:vertical_jeans}. The results are shown in Table~\ref{tab:Galactic_parameters}, where the metallicity bins were selected to include at least 30 stars each. Though we found that both $h_R$ and $h_Z$ increased with decreasing metallicity (except for the most metal-poor bin), this trend is not strong enough  to stand-out significantly from the errors. We conclude that  within the errors, the same scale lengths and scale heights are found, which is the signature of only one population.  
Indeed, if a significant amount of relics of a destroyed massive satellite should exist in our line-of-sight, as suggested by \citet{Gilmore_2002}, 
one would expect them to have a  different spatial distribution compared to the canonical thick disc, which we do not observe. 
Unless, of course, the satellite debris provides the dominant stellar population in the thick disc. 

%This result can also be discussed in the frame of a thick disc formed according to a radial migration scenario. In that case, 
%the older and more metal poor stars at the solar radius, coming from the inner parts of the Galaxy, are expected to have a 
%higher vertical velocity dispersion, and hence, should exhibit  scale heights dependent on metallicity. 
%This trend is not seen in our data (if it exists, it should be rather small), which challenges the migration scenario as being the most important 
%process of creation of the Galactic thick disc. 

This result can also be discussed in the frame of a thick disc formed according to a radial migration scenario. In that case, 
the older stars that are at the solar radius have come from the inner parts of the Galaxy, and are expected to have a 
higher vertical velocity dispersion and a different metallicity, and hence, should exhibit  scale heights dependent on metallicity. 
In particular, the model of \citet{Radial_mixing_2009} predicts a lower scale height for the metal-poor thick disc compared to its 
metal-rich counter part. 
This trend is not observed in our data (if it exists, it should be fairly small), which challenges the migration scenario as being the most important 
process of creation of the galactic thick disc.

\begin{table*}
\centering
\caption{Kinematic parameters, radial scale lengths and scale heights for different metallicity bins of the thick disc targets.}
\label{tab:Galactic_parameters}

\begin{tabular}{cccccccccc}
\hline
\hline

$\overline{[M/H]}$ &     N  & $\overline{V_R}$ & $\sigma_{V_R}$ &  $\overline{V_\phi} $ & $\sigma_{V_\phi}$ & $\overline{V_Z}$ & $\sigma_{V_Z}$ & $h_R$ & $h_Z$\\ 
 (dex) & & (\mbox{km s}$^{-1}$) &  (\mbox{km s}$^{-1}$) &(\mbox{km s}$^{-1}$) &  (\mbox{km s}$^{-1}$) &(\mbox{km s}$^{-1}$) &  (\mbox{km s}$^{-1}$) & (kpc)& (pc) \\\hline

-1.14&          36&    -5$\pm$     9&    58$\pm$    11&  -137$\pm$    11&    61$\pm$     7&    -7$\pm$     8&    59$\pm$     7& 1.9$\pm$ 0.7&   934$\pm$   166\\
 -0.67&          26&    -3$\pm$    17&    85$\pm$    17&  -161$\pm$    11&    54$\pm$    11&    -4$\pm$    12&    54$\pm$     8& 4.0$\pm$ 1.3&   804$\pm$   181\\
 -0.40&          56&     5$\pm$     8&    81$\pm$     8&  -168$\pm$     6&    52$\pm$     5&   -17$\pm$     6&    45$\pm$     4& 3.8$\pm$ 0.9&   610$\pm$    90\\
 -0.11&          37&     6$\pm$     9&    64$\pm$     8&  -171$\pm$     7&    50$\pm$     6&   -18$\pm$     7&    45$\pm$     5& 3.1$\pm$ 0.9&   620$\pm$    97\\

 \hline
\end{tabular}
\tablefoot{All velocity dispersions were corrected for the observational errors. The scale lengths $h_r$ and scale heights $h_z$ were computed using Eq.~\ref{eqn:jeans2}~ and ~\ref{eqn:vertical_jeans}, and we assumed $K_Z=2\pi G \times 71~M_\odot \mathrm{pc}^{-2}$. }
\end{table*}

%
%________________________________________________________________________________________________
\section{Conclusions}
 A significant sample of roughly 700 low-resolution spectra (FLAMES/GIRAFFE LR8 setup, covering the IR \ion{Ca}{ii} triplet) of stars 
faint enough to probe long distances and bright enough to get high signal-to-noise ratios were collected towards the galactic 
coordinates $l\sim 277^{\circ}$, $b\sim 47^{\circ}$. 
The stellar atmospheric parameters ($T_\mathrm{eff}$, $\log~g$, overall metallicity [M/H]) were extracted from the spectra with our 
automated code, which is fully described in \citet{Kordopatis11a}. Given the proper motions of \citet{Ojha_photometry} and our derived 
radial velocities, we were able to derive the distances and the positions for  479 stars of our sample and  the 3D motions 
and the orbital eccentricities for  452 of them.

 We found a thick disc with a mean rotational velocity $\overline{V_\phi}\sim-167$~\mbox{km s}$^{-1}$, a value slightly higher than the commonly adopted lag  of 
less than 50~\mbox{km s}$^{-1}$. We emphasise that our sample of stars probes distances above 1~kpc from the plane, so this difference may 
imply a correlation between lag velocity and vertical velocity rather than a simple inconsistency with local data. 
The mean measured metallicity of the thick disc is $-0.45$~dex. Its metal-poor tails extends to $-1.8$~dex, whereas its metal-rich 
tail goes up to solar and super-solar values. 
Depending on how the thick disc stars are selected ($Z$ positions or velocities), these values may vary a little, but generally agree well between them.  
 The vertical velocity gradient, \VphiZ, and metallicity gradient \MetaZ\ that are measured for 
the regions where the thick disc is the dominant population ($1<Z<4$~kpc) seem to agree well with a smooth transition 
between the galactic components, and are compatible with the values found by \citet{Chiba_and_Beers,Allende_Prieto_SDSS,Girard_2006, Lee11} 
and \citet{Ruchti11}, whose values range from $16\pm4$ to $30\pm3$~\mbox{km s}$^{-1}$\mbox{ kpc}$^{-1}$ and from $-0.07\pm0.01$ to $-0.09\pm0.05$~dex~\mbox{kpc}$^{-1}$. 
A correlation between the metallicity and the rotational velocity of $\partial V_\phi / \partial [M/H]=43 \pm 11$~\mbox{km s}$^{-1}$~\mbox{dex}$^{-1}$ is also found, in agreement with \citet{Spagna_2010} and \citet{Lee11}. 

The radial scale length and scale height of the thick disc are estimated to be $h_R\sim 3.4\pm0.7$~kpc and $h_Z\sim 694 \pm 45$~pc, 
which agrees well with the SDSS results of \citet{Juric_2008}. No clear metallicity dependences are detected for $h_R$ or for $h_Z$, 
pointing towards a thick disc that is mainly composed  by only one population. 

Finally, we found a broad peak of the eccentricity distribution for the thick disc around $\epsilon\sim0.3$ which, according to the 
works of \citet{Sales_2009} and \citet{Di_Matteo_2011}  seem to rule out a pure accretion scenario. These results agree 
with those recently obtained by \citet{Wilson_RAVE_eccentricities} and \citet{Eccentricities_SDSS},  who  measured the eccentricity 
distributions for a sample of RAVE and SEGUE stars.

However, several questions still remain open. For instance, we found difficulties in fitting the transition between the thin and the 
thick disc simply by adjusting parameters in the Besan\c{c}on model. In addition, the plateau for the high-metallicity and low-altitude 
stars ({\it i.e.} the thin disc) seems to suggest a local density of the thick disc  around 18\%, higher than that assumed by that model. 
Finally, the existence of intrinsic vertical gradients in the thick disc cannot be ruled out, because we did not obtain a sufficiently 
well defined (and statistically large enough) sample of thick disc members. 
Additional targets to increase the statistics and higher resolution spectra on well selected spectral domains to separate thin disc stars from the thick disc ones with chemical content criteria are hence strongly recommended for future surveys.     

%Angular momentum appears too low to arise from dynamical
%perturbations induced by the galactic bar and coincides with that of the kinematically peculiar population of stars
%identified above and below the galactic plane by Gilmore,

%The X coordinate varies. Are the seen effects induced by a vertical change in the properties, or a radial one? Or both ? 
%De Jong 2010, Ivezic 2008 sur la overdensity

\onltab{1}{
\begin{table*}
\caption{Kinematics of the selected targets belonging to the final catalogue}
\label{online3}
\scriptsize
\begin{tabular}{ccccccccccccccccc}
ID &  $\mu_l$     & $\mu_b$    & $V_{\rm rad}$ & $\Delta_{V_{\rm rad}}$ & U              & $\Delta_{U}$     & V              & $\Delta_{V}$     & W              & $\Delta_{W}$     & $V_R $ &$\Delta_{V_R}$ & $V_\phi $ &$\Delta_{V_\phi}$ &$\epsilon$ & $\Delta_{\epsilon}$    \\
   &  ($"/cen$)   & ($"/cen$) & (km/s) & (km/s)          & (km/s)  &   (km/s)  &  (km/s)  &   (km/s)  & (km/s)  &   (km/s)  &   (km/s)  &   (km/s)  & (km/s)  &   (km/s)  &  \\ \hline
    1  &   0.142  &  -0.361  &   110.2  &     3.8  &    53  &    50  &  -125  &    41  &    27  &    39  &   -90  &    50  &   -66  &    55  &    0.54  &    0.16 \\
    2  &   0.382  &   0.107  &   141.3  &     5.3  &   132  &    68  &   -47  &    48  &   127  &    43  &  -212  &    69  &   -73  &    56  &    0.59  &    0.14 \\
    3  &   0.361  &  -0.263  &     9.0  &     4.5  &     8  &     3  &    -8  &     4  &     3  &     4  &   -25  &     3  &  -216  &     4  &    0.06  &    0.01 \\
    4  &   0.354  &  -0.739  &    24.7  &     6.4  &   114  &    51  &  -134  &    42  &  -106  &    39  &  -149  &    49  &   -28  &    48  &    0.44  &    0.08 \\
...
\end{tabular}
\normalsize
\tablefoot{$\mu_l$ and $\mu_b$ are taken from \citet{Ojha_photometry}. $U$, $V$ and $W$ are the cartesian velocity coordinates, with respect to the LSR, hence, without taking into account the peculiar solar velocities. $V_R$ and $V_\phi$ are the velocity components in cylindrical coordinates, centred at the galactic centre. }
\end{table*}
} % endof online table

\onltab{2}{
\begin{table*}
\caption{Individual atmospheric parameters (as derived by the procedure of Paper~I), relative V-Magnitudes, (B-V) colour, absolute V-magnitude and estimated S/N of the observed targets of the final catalogue}
\label{online}
\begin{tabular}{ccccccccccccccccccccccccc}
ID & $T_\mathrm{eff}$  & $\Delta_{T_\mathrm{eff}}$ & $\log~g$  & $\Delta_{\log~g}$ & [M/H]    & $\Delta_{\rm [M/H] }$ &$m_V$ & (B-V) & $M_V$ & $\Delta_{M_V}$ &S/N \\ 
   &  (K)    & (K)          & (cm~s$^{-2})$& (cm~s$^{-2})$  & (dex)   & (dex)            &      &       &       &                &(pixel$^{-1}$) \\ \hline
    1  &  5962  &   212  &    4.02  &    0.30  &   -0.76  &    0.15  &   17.37  &    0.61  &    3.98  &    0.86  &   23 \\
    2  &  5001  &    74  &    3.51  &    0.12  &   -0.25  &    0.10  &   17.24  &    0.69  &    3.01  &    0.41  &   29 \\
    3  &  4272  &    64  &    4.86  &    0.12  &   -0.22  &    0.09  &   16.41  &    1.42  &    8.30  &    0.13  &   60 \\
    4  &  4997  &    88  &    3.73  &    0.17  &   -1.00  &    0.16  &   16.81  &    0.73  &    3.14  &    0.29  &   35 \\
...

\end{tabular}
\tablefoot{$m_v$ and (B-V) are taken from \citet{Ojha_photometry}.}
\end{table*}
} % endof online table

\onltab{4}{
\begin{table*}
\caption{Positions of the selected targets belonging to the final catalogue}
\label{online2}
\begin{tabular}{ccccccccccccc}
%ID & $l$& $b$& $V_\mathrm{rad}$&$\Delta V_\mathrm{rad}$& $\mu_l$ & $\mu_b$&  $D$ & $\Delta D$&  $X$ & $\Delta X$&  $Y$ & $\Delta Y$ &  $Z$ & $\Delta Z$ &$U$ & $\Delta U$  & $V$ & $\Delta V$ &   $W$ & $\Delta W$  

ID &  $l$         & $b$        & D    & $\Delta_{D}$ & X    & $\Delta_{X}$ & Y    & $\Delta_{Y}$ & Z    & $\Delta_{Z}$ &R   & $\Delta_{R}$ \\ 
   &  ($^\circ$)  & ($^\circ$) & (pc) & (pc)        & (pc) & (pc)         & (pc) &  (pc)       & (pc) & (pc)         &(pc)&  (pc) \\ \hline
    1  & 280.201  &  47.452  &      4573  &      2061  &       552  &       245  &     -3069  &      1365  &      3397  &      1511  &      8167  &       326 \\
    2  & 280.338  &  47.363  &      6725  &      1333  &       816  &       159  &     -4477  &       876  &      4942  &       967  &      8504  &       329 \\
    3  & 280.308  &  47.406  &       401  &        24  &        48  &         3  &      -267  &        16  &       295  &        18  &      7955  &         2 \\
    4  & 280.223  &  47.409  &      5202  &       705  &       628  &        84  &     -3486  &       467  &      3853  &       516  &      8167  &       124 \\

...
\end{tabular}
\tablefoot{$l$ and $b$ are the galactic coordinates taken from \citet{Ojha_photometry}. $D$ is the line-of-sight distance. $X$, $Y$and $Z$ are heliocentric distances in a cartesian reference frame. $R$ is the galactocentric planar radial coordinate. }
\end{table*}
} % endof online table

\begin{acknowledgements} 
The authors would like to thank the MESOCENTRE de l'Observatoire de la C\^{o}te d'Azur for computing facilities. We are grateful to M.~Irwin for letting us use the routine of sky subtraction, A.~Robin for her useful advice on the use of the Besan\c{c}on model and the referee for his useful comments that improved the quality of our article. Finally, G.K. would like to thank the Centre National d'Etudes Spatiales (CNES) and the Centre National de Recherche Scientifique (CNRS) for the financial support.  
MZ acknowledges Proyecto Fondecyt Regular \#1110393, The Fondap
Center for Astrophysics 15010003, BASAL CATA PFB-06, and Programa
Iniciativa Cientifica Milenio, awarded to The Milky Way Millennium
Nucleus P07-021-F.
\end{acknowledgements}

\bibliographystyle{aa} % style aa.bst
\bibliography{kordopatis} % your references Yourfile.bib

\end{document}